\documentclass[aps,prb,twocolumn,longbibliography,superscriptaddress]{revtex4-1}

\usepackage{graphicx}
\usepackage{epstopdf}
\usepackage{amsmath}
\usepackage{amsfonts}
\usepackage{amssymb}
\usepackage{hyperref}
\usepackage{bm}
\usepackage{makecell}

\usepackage{hyperref}

\usepackage{graphicx}
\usepackage{dcolumn}
\usepackage{bm}
\usepackage{color}

\begin{document}



\title{Nematicity of correlated systems driven by anisotropic chemical phase separation}

\author{Ye Yuan}
 \affiliation{Helmholtz-Zentrum Dresden-Rossendorf, Institute of Ion Beam Physics and Materials Research,
 Bautzner Landstrasse 400, D-01328 Dresden, Germany}

\author{Ren\'e H\"ubner}
 \affiliation{Helmholtz-Zentrum Dresden-Rossendorf, Institute of Ion Beam Physics and Materials Research,
 Bautzner Landstrasse 400, D-01328 Dresden, Germany}

\author{Magdalena Birowska}
 \affiliation{Faculty of Physics, Institute of Theoretical Physics, University of Warsaw, Pasteura 5, PL-02093 Warsaw, Poland}

\author{Chi Xu}
 \affiliation{Helmholtz-Zentrum Dresden-Rossendorf, Institute of Ion Beam Physics and Materials Research,
 Bautzner Landstrasse 400, D-01328 Dresden, Germany}
 \affiliation{Technische Universit\"at Dresden, D-01062 Dresden, Germany}

 \author{Mao Wang}
 \affiliation{Helmholtz-Zentrum Dresden-Rossendorf, Institute of Ion Beam Physics and Materials Research,
 Bautzner Landstrasse 400, D-01328 Dresden, Germany}
 \affiliation{Technische Universit\"at Dresden, D-01062 Dresden, Germany}

 \author{Slawomir Prucnal}
 \affiliation{Helmholtz-Zentrum Dresden-Rossendorf, Institute of Ion Beam Physics and Materials Research,
 Bautzner Landstrasse 400, D-01328 Dresden, Germany}

 \author{Rafal Jakiela}
 \affiliation{Institute of Physics, Polish Academy of Sciences,
Aleja Lotnikow 32/46, PL-02668 Warsaw, Poland}

\author{Kay Potzger}
 \affiliation{Helmholtz-Zentrum Dresden-Rossendorf, Institute of Ion Beam Physics and Materials Research,
 Bautzner Landstrasse 400, D-01328 Dresden, Germany}

 \author{Roman B\"ottger}
 \affiliation{Helmholtz-Zentrum Dresden-Rossendorf, Institute of Ion Beam Physics and Materials Research,
 Bautzner Landstrasse 400, D-01328 Dresden, Germany}

\author{Stefan Facsko}
 \affiliation{Helmholtz-Zentrum Dresden-Rossendorf, Institute of Ion Beam Physics and Materials Research,
 Bautzner Landstrasse 400, D-01328 Dresden, Germany}

\author{Jacek A. Majewski}
 \affiliation{Faculty of Physics, Institute of Theoretical Physics, University of Warsaw, Pasteura 5, PL-02093 Warsaw, Poland}

\author{Manfred Helm}
 \affiliation{Helmholtz-Zentrum Dresden-Rossendorf, Institute of Ion Beam Physics and Materials Research,
 Bautzner Landstrasse 400, D-01328 Dresden, Germany}
 \affiliation{Technische Universit\"at Dresden, D-01062 Dresden, Germany}

\author{Maciej Sawicki}
 \affiliation{Institute of Physics, Polish Academy of Sciences,
Aleja Lotnikow 32/46, PL-02668 Warsaw, Poland}

\author{Shengqiang Zhou} \email{s.zhou@hzdr.de}
 \affiliation{Helmholtz-Zentrum Dresden-Rossendorf, Institute of Ion Beam Physics and Materials Research,
 Bautzner Landstrasse 400, D-01328 Dresden, Germany}

 \author{Tomasz Dietl} \email{dietl@MagTop.ifpan.edu.pl}
\affiliation{International Research Centre MagTop, Institute of Physics, Polish Academy of Sciences,
Aleja Lotnikow 32/46, PL-02668 Warsaw, Poland}
\affiliation{WPI-Advanced Institute for Materials Research, Tohoku University, Sendai 980-8577, Japan}

\date{\today}

\begin{abstract}
The origin of nematicity, i.e., in-plane rotational symmetry breaking, and in particular the relative role played by spontaneous unidirectional ordering of spin, orbital, or charge degrees of freedom, is a challenging issue of magnetism, unconventional superconductivity, and quantum Hall effect systems, discussed in the context of doped semiconductor systems, such as Ga$_{1-x}$Mn$_x$As, Cu$_x$Bi$_2$Se$_3$, and Ga(Al)As/Al$_x$Ga$_{1-x}$As  quantum wells, respectively. Here, guided by our experimental and theoretical results for In$_{1-x}$Fe$_x$As, we demonstrate that spinodal phase separation at the growth surface (that has a lower symmetry than the bulk) can lead to a quenched nematic order of alloy components, which then governs low temperature magnetic and magnetotransport properties, in particular the magnetoresistance anisotropy whose theory for the $C_{2v}$ symmetry group is advanced here. These findings, together with earlier data for Ga$_{1-x}$Mn$_x$As, show under which conditions anisotropic chemical phase separation accounts for the magnitude of transition temperature to a collective phase or merely breaks its rotational symmetry. We address the question to what extent the directional distribution of impurities or alloy components setting in during the growth may account for the observed nematicity in other classes of correlated systems.

\end{abstract}

\maketitle



\section{Introduction}

As noted by Kramers \cite{Kramers:1934_P}, quantum electron hopping between anion $p$ states and open $d$ orbitals of magnetic cations, i.e., $p$--$d$ hybridization, results in exchange interactions between localized spins. An intricate character of this coupling mechanism accounts for the richness of spontaneous spin and orbital orderings \cite{Tokura:2000_S} as well as for classical and quantum spin dynamics \cite{Zhou:2017_RMP}, the questions extensively studied over past decades in the abundant class of magnetic and superconducting compounds. It is, for instance, presently being discussed whether charge, spin, or orbital spatial correlations account for nematic characteristics of iron-based superconductors \cite{Fernandes:2014_NP,Bohmer:2018_JPCM} or superconductivity in doped topological insulators \cite{Behnia:2017_NP}. However, it has also been realized that $p$--$d$ hybridization leads to attractive forces between transition metal (TM) cations in semiconductors \cite{Schilfgaarde:2001_PRB,Sato:2005_JJAP}. It becomes increasingly clear that these forces generate various patterns of the {\em TM distribution} in dilute magnetic semiconductors, predefined by growth and processing conditions \cite{Dietl:2015_RMP}. A question then arises about the role played by quenched non-randomness in the nematicity of collective phases, i.e, in the unexpected two-fold in-plane anisotropy of tetragonal or trigonal ferromagnets and superconductors.

Recent comprehensive studies of (In,Fe)As and (In,Fe)As:Be grown by low-temperature molecular beam epitaxy (LT-MBE) \cite{Hai:2012_APL_A,Hai:2012_APL_B,Sakamoto:2016_PRB} have indicated that this system forms a class of materials with properties distinctly different compared to those found for (Ga,Mn)As and (Ga,Mn)As:Be in which holes mediate ferromagnetic interactions between randomly distributed Mn ions \cite{Dietl:2014_RMP,Jungwirth:2014_RMP}. In particular: (i) (In,Fe)As:Be is $n$-type but nevertheless ferromagnetic \cite{Hai:2012_APL_A}; (ii) the shape component dominates magnetic anisotropy \cite{Hai:2012_APL_A}, whereas the crystalline contribution, breaking rotational symmetry,  governs in the case of (Ga,Mn)As \cite{Dietl:2014_RMP}; (iii) the anisotropic magnetoresistance (AMR) depends on the orientation of magnetization with respect to crystallographic axes (the "crystalline" AMR) \cite{Hai:2012_APL_B} rather than on the current direction as in (Ga,Mn)As in which the "noncrystalline" AMR dominates \cite{Rushforth:2007_PRL}; and (iv) in contrast to (Ga,Mn)As and (Ga,Mn)As:Be, the distribution of Fe ions is nonuniform in (In,Fe)As:Be \cite{Sakamoto:2016_PRB}.

In this paper we report on a series of experiments carried out for recrystallized films of Fe-implanted InAs, which reveals a hitherto unobserved character of the distribution of magnetic ions in a semiconductor host and the associated magnetic properties. With the aid of {\em ab initio} computations we identify microscopic mechanisms accounting for the observed anisotropic nanoscale chemical phase separation, and we explain surprising properties of magnetic anisotropy and AMR, making use of the AMR theory developed here for the $C_{2v}$ symmetry group up to eighth order. In more general terms, we discuss how the sign and magnitude of the charge transfer energy specifying a particular magnetic impurity in a given host \cite{Mizokawa:2000_PRB,Dietl:2002_SST} determine not only the electrical activity of the impurity and its charge state, but also the incorporation of non-magnetic dopants and the aggregation of the magnetic constituent, contrasting (In,Fe)As and (Ga,Mn)As. Our results indicate  that in specific cases, quenched anisotropic distribution of dopants or defects rather than spontaneous in-plane uniaxial ordering of electronic degrees of freedom can account, often via spin-orbit interactions, for nematicity of low-temperature collective phases. In this context we refer---in the final part of our paper---to the case of intercalated superconducting FeSe and Bi$_2$Se$_3$. We also note that AMR theory discussed here may elucidate the origin of "noncrystalline" in-plane AMR of Ga(Al)As/Al$_x$Ga$_{1-x}$As  quantum wells in the quantum Hall effect regime, whereas the lowering of symmetry to $C_{2v}$ in these epitaxial structures could explain crystalline anisotropy of longitudinal resistance observed under these conditions.

\section{Visualization of anisotropic spinodal phase separation in recrystallized $\mbox{(In,Fe)As}$}
\subsection{Growth of (In,Fe)As layers}
Implantation of Fe ions into a (001) InAs wafer is carried out at an energy of 100\,keV to a fluence of $1\times10^{16}$\,cm$^{-2}$. During implantation, the wafer is tilted by $7^{\circ}$ with respect to the ion beam to avoid channeling effects. According to the stopping and range of ions in matter (SRIM) simulation code \cite{Ziegler:2004_NIMB}, the projected range ($R_P$) and the longitudinal straggling ($\Delta R_P$) for the Fe distribution are calculated to be 58 and 36\,nm, respectively. Then, a UV pulsed laser (pulse duration = 28\,ns) with a wavelength of 308\,nm is employed to recrystallize the as-implanted InAs layer \cite{Scarpulla:2003_APL,Scarpulla:2005_PRL}. During the pulsed laser melting (PLM) process, the near-surface layer including the whole Fe-implanted region is molten, whereas the bulk substrate remains at ambient temperature. After the pulse, the molten layer starts to cool down. The resulting recrystallization process, proceeding from the interface to the surface, completes within a microsecond time range.

\subsection{Structural and chemical characterization}
\subsubsection{Electron microscopy and energy-dispersive x-ray spectroscopy}
\begin{figure}[tb]
\includegraphics[width=8.5 cm]{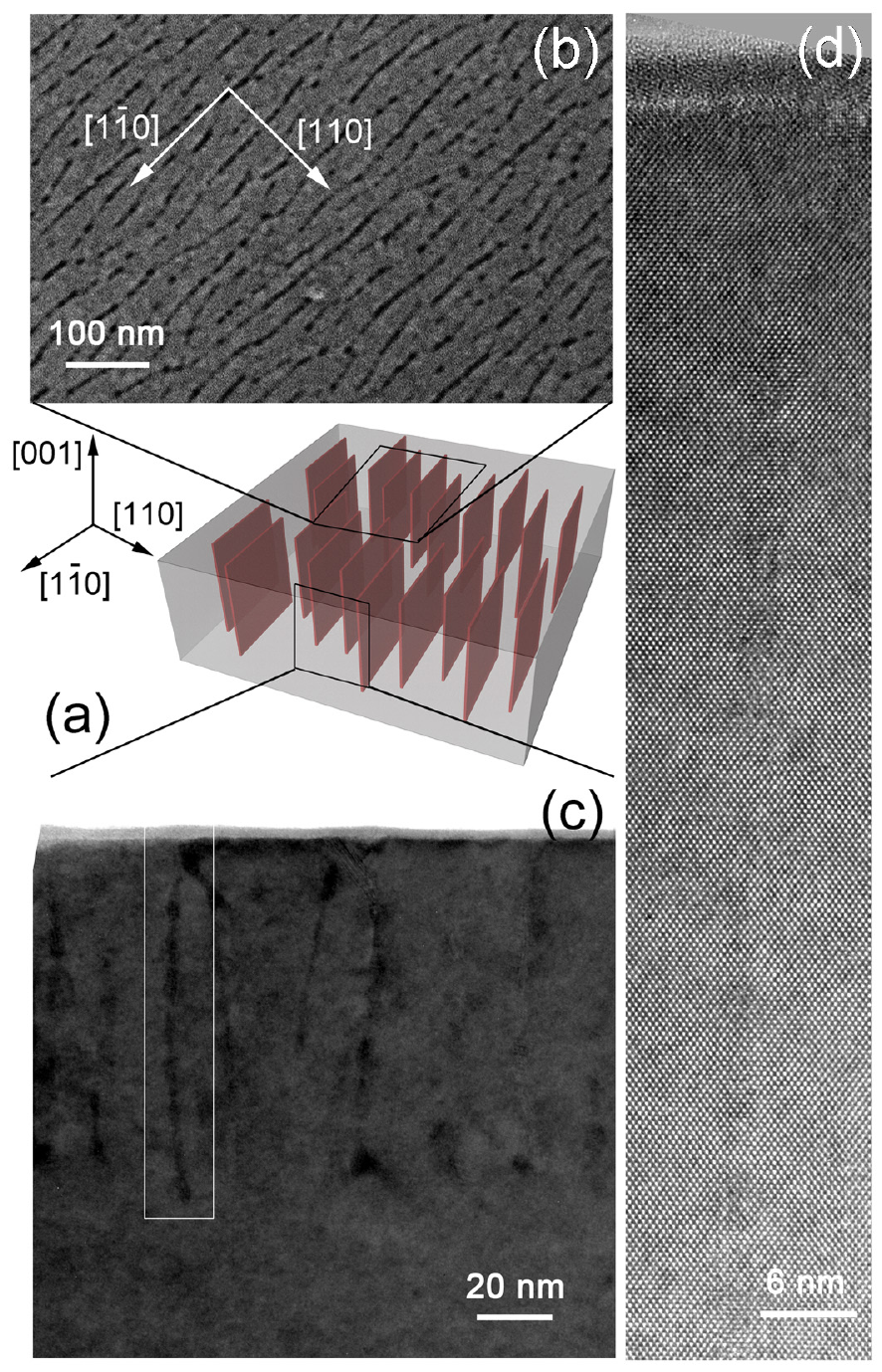}
 \caption{\label{fig:TEM} (Color online) The morphology of Fe-rich nano-lamellae in single-crystalline InAs. (a) Schematic image of the nano-lamellae and their orientation within the InAs matrix. Structural characterization by (b) top-view scanning electron microscopy (SEM), (c) cross-sectional bright-field transmission electron microscopy (TEM) in $[1\bar{1}0]$ zone axis geometry, and (d) high-resolution TEM of the area marked with the white rectangle in part (c) point to the pseudomorphic growth of nano-lamellae oriented in (110) planes of the single-crystalline (001) InAs wafer.}%
\end{figure}

\begin{figure}[tb]
\includegraphics[width=8.5 cm]{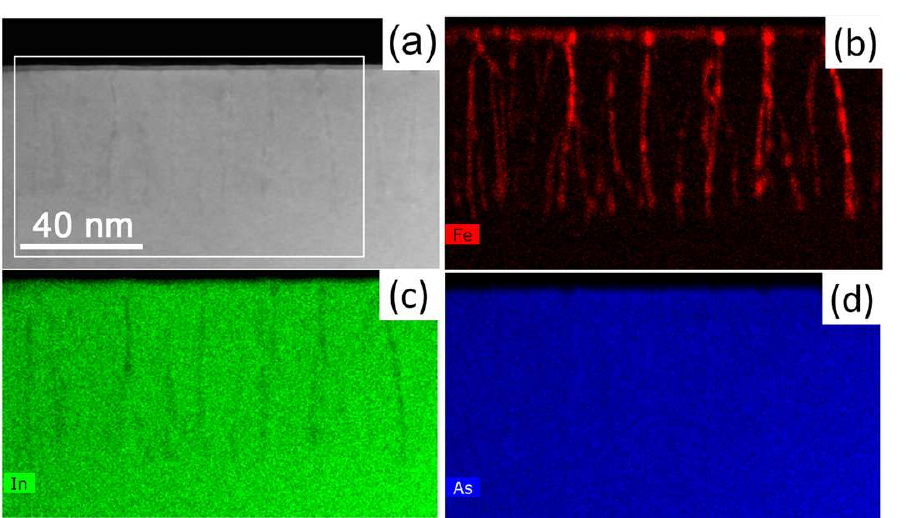}
 \caption{\label{fig:EDXS} (Color online) Chemical analysis of the nano-lamellar structure. (a) High-angle annular dark-field scanning TEM image of the same specimen region as shown in the bright-field TEM micrograph of Fig.~\ref{fig:TEM}(c) together with the (b) iron, (c) indium, and (d) arsenic element distributions obtained by EDXS for the area marked with the white rectangle in part (a). }%
\end{figure}

The results of structural and chemical nanocharacterization by electron microscopy of our (In,Fe)As layers, collected in Figs.~\ref{fig:TEM} and \ref{fig:EDXS}, reveal the presence of a few nanometers' thick and about 90\,nm long Fe-rich (Fe,In)As nanocrystals embedded in the InAs matrix in the form of lamellae parallel to the (110) planes. These nanocrystals are only observed in the PLM-regrown layer. The aggregation of Fe cations has a character of the chemical phase separation as it occurs without deteriorating the zinc-blende crystal structure [Fig.~\ref{fig:TEM}(d)] and, it can therefore, be determined only by element-specific methods. This is further confirmed by energy-dispersive x-ray spectroscopy (EDXS) showing the uniform As distribution, while there is Fe in the regions where In gets depleted (Fig.~\ref{fig:EDXS}).

\subsubsection{Secondary ion mass spectroscopy and Rutherford backscattering}
\label{sec:SIMS}
\begin{figure}
\includegraphics[width=7.5 cm]{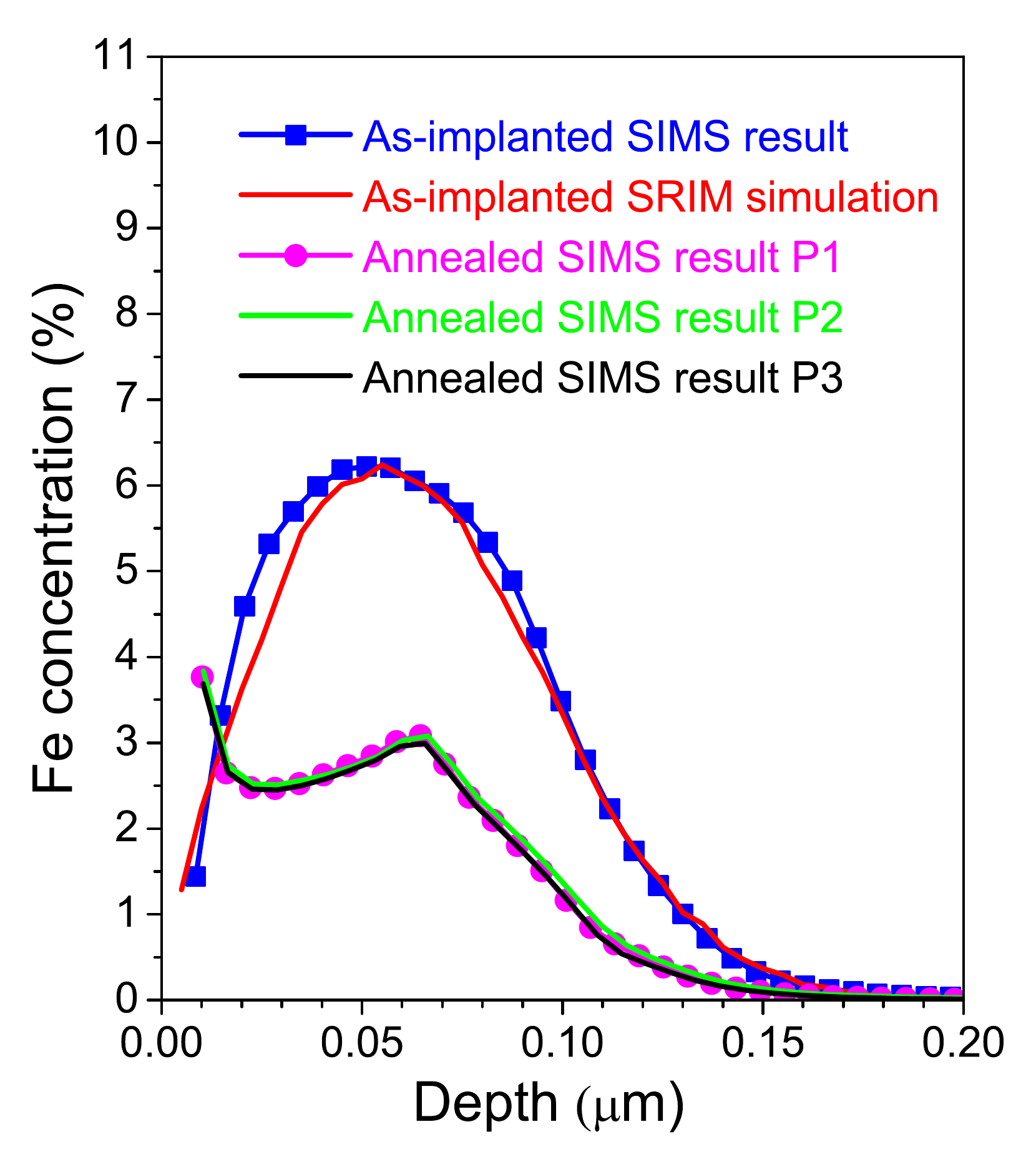}
 \caption{\label{fig:SIMS} Depth dependent Fe concentration by secondary ion mass spectrometry (SIMS) and SRIM simulation.}
\end{figure}

Secondary ion mass spectrometry (SIMS) measurements have been employed to obtain information on the Fe depth profile of as-implanted and laser treated samples. Three different sample regions have been chosen to check layer homogeneity (Fig.~\ref{fig:SIMS}). Iron atoms reside in a range from the surface down to a depth of 150\,nm. An overlap of curves showing results for three scans proves the homogeneity of the film at a micrometer scale. It is worth mentioning that after pulsed laser annealing almost two thirds of implanted atoms diffuse into the surface and form an amorphous Fe-rich layer, which gives rise to the intense Fe SIMS signal at the surface region. The remaining Fe atom density in InAs lies between 2.5\% and a maximum value of 3.1\% within the longitudinal straggling region. The shape of the iron atom distribution is  modified by the laser pulse: the Gaussian shape disappears and the peak position shifts to around 70\,nm.

\begin{figure}
\includegraphics[width=8.5 cm]{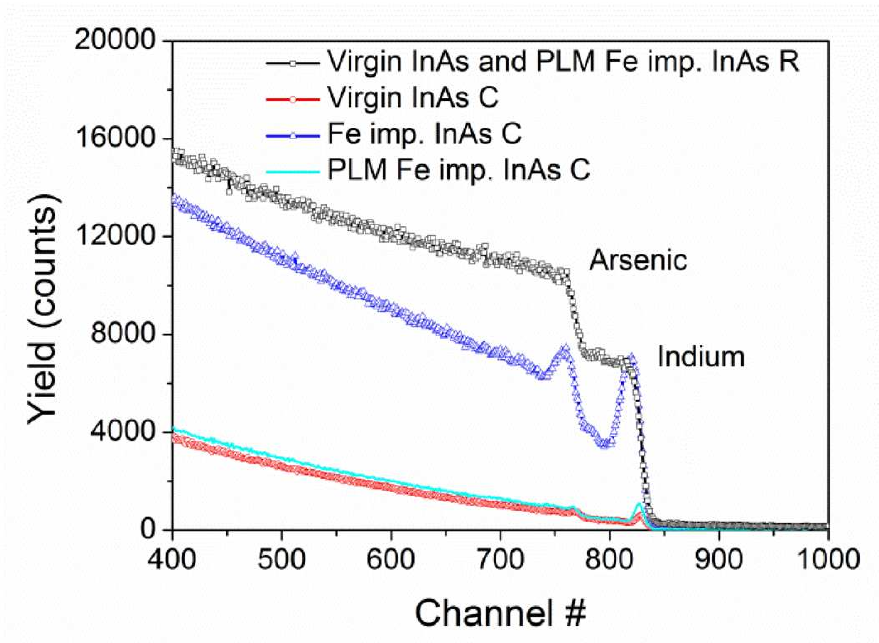}
 \caption{\label{fig:RBS} Rutherford backscattering spectrometry random (R) and channeling (C) spectra for virgin InAs (squares and circles, respectively), an as-Fe-implanted InAs (triangles) and pulsed laser melting treated (solid lines) samples.}
\end{figure}

The recrystallization of the Fe-rich nano-region and of the InAs matrix is investigated by Rutherford backscattering spectrometry (RBS)/channeling spectrometry. In particular, the channeling effect appears if the film is fully recrystallized and pseudomorphic with the InAs substrate. During the RBS measurement, a collimated 1.7\,MeV He$^+$ beam with a 10-20\,nA beam current is applied, and the scattered ions are collected at a backscattering angle of 170$^{\circ}$. The channeling spectra are recorded by aligning the InAs [001] axis parallel to the impinging He$^+$ beam. The selected RBS spectra are plotted in Fig.~\ref{fig:RBS} allowing a comparison of the crystallization degree of a virgin InAs, an as-Fe-implanted InAs without any treatment, and after melting by the pulsed laser and subsequent recrystallization. From the random spectra, indium and arsenic signals are both prominent, whereas the Fe signal is not visible due to its low concentration of only several percent and overlapping with the arsenic and indium signals. As shown in Fig.~\ref{fig:RBS}, the channeling effect in the as-implanted layer is strongly suppressed, similarly to Mn-implanted GaAs \cite{Burger:2010_PRB}, indicating that the Fe-implantation leads to sizable damage of the InAs matrix. However, after PLM, the presence of the channeling effect confirms the recrystallization of the implanted layer and an incorporation of Fe atoms onto lattice sites \cite{Benzeggouta:2014_PRB}.

Interestingly, the Fe-doped layer quality after regrowth compares favorably to the quality of the virgin InAs wafer. To quantify the crystalline quality, $\chi_{\text{min}}$, the ratio of the backscattering yield between the channeling and the random spectra, is calculated. Values of 7\% and 5.4\% are obtained for the PLM regrown Fe doped InAs and the reference InAs substrate, respectively. This crystalline quality is comparable to homogeneous epitaxial films of dilute magnetic semiconductors (DMSs) prepared by the same approach, e.g. (Ga,Mn)As and (Ga,Mn)P \cite{Scarpulla:2003_APL,Dubon:2006_PB}. Importantly, $\chi_{\text{min}}$ for our Fe-implanted and subsequent recrystallized InAs samples is significantly smaller compared to Fe implanted ZnO or TiO$_2$ after long time furnace annealing, where the crystalline phase separation (i.e., precipitation of a secondary phase) takes place \cite{Zhou:2008_JAP,Potzger:2006_APL}. Altogether, the results of structural and chemical characterization give a strong evidence for a single-crystalline structure of the PLM treated Fe-implanted InAs samples.

In summary, the nano-lamellae length determined by TEM matches the Fe distribution width measured by SIMS.
In addition, according to the Rutherford backscattering spectrometry, no interstitial atoms are present in the Fe-doped region.

\subsubsection{Conversion electron M\"ossbauer spectroscopy}
\begin{figure}
\includegraphics[width=8.5 cm]{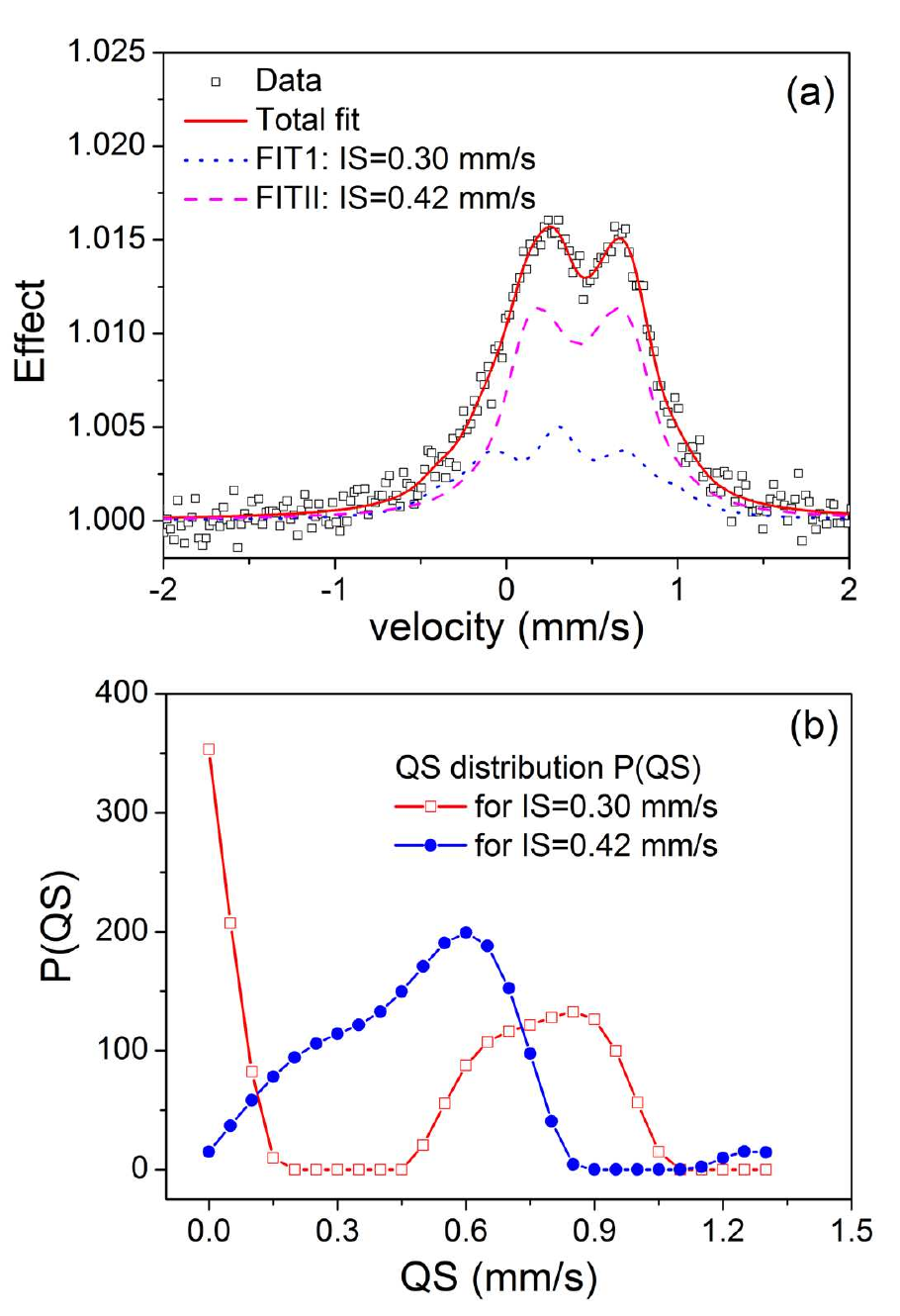}
 \caption{\label{fig:CEMS} (a) Conversion electron M\"ossbauer spectroscopy (CEMS) spectrum recorded at room temperature fitted with two quadrupole-split emission lines exhibiting a broad distribution of the quadrupole splittings. Parameters can be found in the text. (b)  Behavior of the quadrupole-splitting for both distributions.}
\end{figure}

One recrystallized film has also been probed by means of room-temperature conversion electron M{\"{o}}ssbauer spectroscopy (CEMS). The spectrum [Fig.~\ref{fig:CEMS}(a)] shows a broad asymmetric doublet that points to a distribution of the hyperfine parameters arising from the quadrupole interaction. Using the NORMOS routine \cite{Brand:1987_NIMB} the best fit is achieved assuming two quadrupole distributions shown in Fig.~\ref{fig:CEMS}(b). One of them (68\% of the spectral area) has an isomeric shift of $IS = 0.42$\,mm/s with respect to $\alpha$-Fe and a quadrupole splitting of $QS \sim 0.5$\,mm/s at the maximum of the distribution. The other one (32\% of the spectral area) shows both a single line part with zero splitting and broader splitting between 0.65 and 0.9\,mm/s with an isomeric shift of 0.30 mm/s.

The hyperfine parameters of the first distribution are close to those of FeAs \cite{Haggstrom:1989_EPL}, where $IS = 0.49$\,mm/s and $QS = 0.55$\,mm/s, as well as Fe doped InAs in which $IS = 0.5$\,mm/s and $QS = 0.45$\,mm/s are observed. The higher $s$-electron density and the variation of interatomic distances might be related to the distribution of indium atoms in the neighborhood of the $^{57}$Fe isotope. Our and previous literature results  point to Fe-vacancy complexes in the defective surface region and substitutional Fe atoms in the bulk region. Our experiments are surface sensitive with the probed depth of about 50\,nm. Further hints towards a highly defective surrounding can be found in Ref.~\onlinecite{Bharuth-Ram:2012_NIMB} reporting on the implantation of $^{57}$Mn isotopes  into InAs at 60\,keV at low fluences. The authors found a prominent singlet line that they relate to the substitutional Mn on the indium sites.  The relative contribution of the singlet line increases after annealing while the participation of the doublets ($IS = 0.62$\,mm/s, $QS = 1.22$\,mm/s and $IS =-0.29$\,mm/s, $QS = 0.65$\,mm/s) decreases. The doublet was assigned to the $^{57}$Fe isotopes in a highly defective region such as the surface Fe-rich amorphous layer. Similar observations were reported for Fe doped p-type GaAs \cite{Seregin:2003_Sem}. It was found that in the defective surface region an asymmetric doublet can be observed, which can be decomposed into a doublet and a singlet, the former with $IS = 0.45$\,mm/s and $QS = 0.95$\,mm/s.

In summary, our experiments are surface sensitive with a probed depth of about 50\,nm. In this region, about 70\% of Fe occupy cation substitutional positions in the zinc-blende lattice, i.e., they form Fe--As bonds. The remaining Fe ions reside presumably in the defective and amorphous surface layer.

\section{Magnetic properties of recrystallized  $\mbox{(In,Fe)As}$}

\begin{figure*}[tb]
\includegraphics[width=17.5 cm]{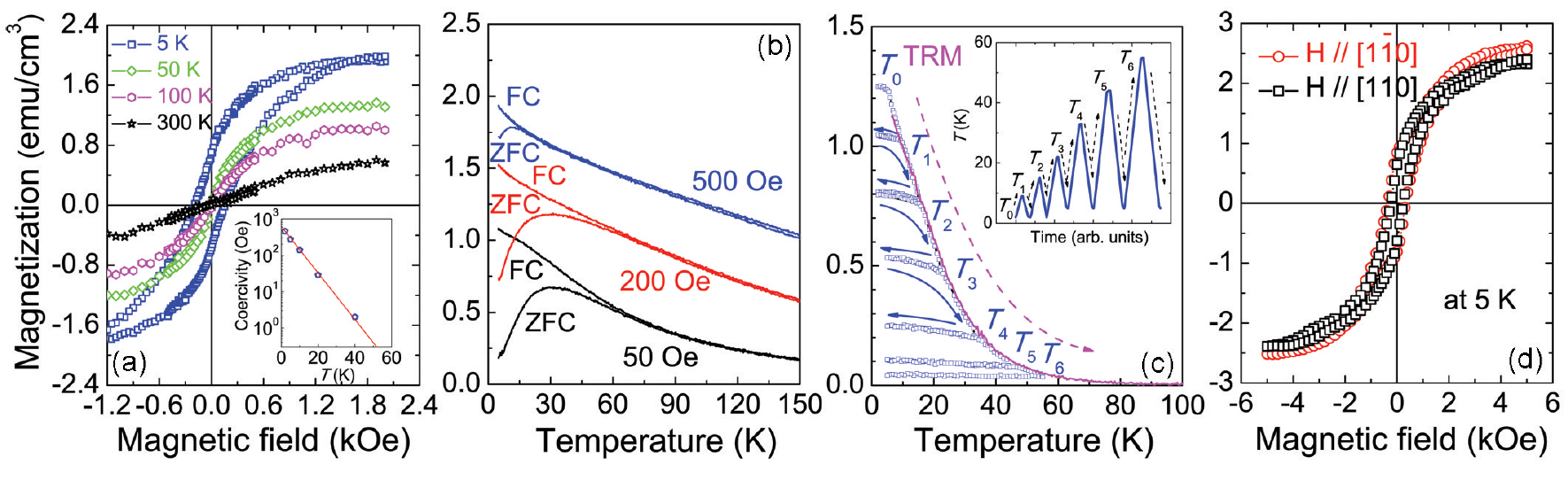}
 \caption{\label{fig:SQUID} (Color online) Magnetic properties of (In,Fe)As epilayers. (a) Magnetic field dependent magnetization measured at temperatures of 5, 50, 100, and 300\,K. The inset shows temperature dependence of the coercivity indicating its exponentially fast decay on temperature. (b) Field cooled (FC) and zero field cooled (ZFC) temperature dependence of magnetization measured at 50, 200, and 500 Oe. (c) Solid line: thermoremanent magnetization (TRM) and circles: its temperature cycling between the base temperature ($2 < T_0 < 5$\,K) and progressively higher temperatures ($T_1$ to $T_6$) according to the pattern drawn in the inset; (d) magnetization loops for the magnetic field along $[1{\bar{1}}0]$ and [110] crystal axes.}%
\end{figure*}

All magnetic measurements are carried out using about ~20\,cm long and 1.5\,mm wide silicon strips to support the investigated specimen in the magnetometer's sample chamber.
The adequate experimental code for minute signals measurements \cite{Sawicki:2011_SST} has been strictly observed.
Importantly for such studies, the truly near-zero field conditions in the sample chamber ($H \simeq 0.1$\,Oe, as established using a Dy$_2$O$_3$ paramagnetic salt) are achieved by degaussing the magnetometer with an oscillating magnetic field of decreasing amplitude from 30\,kOe to about 600\,Oe, followed by a soft quench of the SQUID's superconducting magnet (using the so called "magnet reset" option).
The soft quench is also routinely performed prior to the zero-field studies such as the  thermoremanent moment (TRM, the measurement of the remanent moment on increasing $T$) and during thermal cycling of the sample brought beforehand to its remanence.

Magnetic-field-dependent hysteresis loops at various temperatures of the Fe-implanted InAs sample, studied according to an experimental procedure developed previously \cite{Sawicki:2010_NP,Sawicki:2011_SST}, are shown in Fig.~\ref{fig:SQUID}(a). The magnitude of magnetization and magnetic hysteresis loops point to ferromagnetic coupling within Fe-rich nano-lamellae, which is challenging, particularly considering antiferromagnetic spin ordering found experimentally for FeAs layers in BaFe$_2$As$_2$-type systems \cite{Huang:2008_PRL} and predicted computationally for zinc-blende FeAs \cite{Wei:2003_CTP,Rahman:2006_JMMM}. The coercivity vanishes around 50\,K, decaying exponentially to zero upon increasing temperature, as shown in the inset to Fig.~\ref{fig:SQUID}(a). Although magnetization at a given magnetic field $H$ decreases with temperature, non-zero field-induced magnetization is still detectable at room temperature. This indicates, assuming that a contribution from the Fe-rich amorphous surface layer is negligible, that the Curie temperature $T_{\text{C}}$ of (Fe,In)As is higher than room temperature.

The coercivity behavior, together with the temperature dependence of magnetization after zero-field cooling and field-cooling processes under different fields and temperature cycling presented in Figs.~\ref{fig:SQUID}(b,c), point clearly to the absence of long-range ferromagnetism and a blocked-superparamagnetic (B-SP) character of the studied sample. In particular, each thermal cycle (the whole process is shown in the inset to Fig.~\ref{fig:SQUID}(c)] consists of warming up the sample to a progressively higher temperature followed by re-cooling to the base temperature of below 5\,K. This measurement allows us to distinguish a decaying (with temperature) part of the sample remanent moment (that is the dynamically blocked one by energy barriers) from that related to the spontaneous magnetization in the equilibrium state under  zero field conditions.
This behavior results from the highly non-random distribution of Fe cations, in contrast to the case of (Ga,Mn)As grown by LT-MBE, in which typically weak B-SP signatures coexist with the long-range ferromagnetic order, and stem from the electronic phase separation, i.e., disorder-induced mesoscopic fluctuations in the hole density  \cite{Yuan:2017_PRM,Gluba:2018_PRB}. Interestingly, a tiny asymmetry in the occupation by Mn of the $[1{\bar{1}}0]$  and $[110]$ directions explains a strong uniaxial in-plane magnetic anisotropy of ferromagnetism in (001)(Ga,Mn)As epilayers \cite{Birowska:2012_PRL}. No such in-plane uniaxial magnetic anisotropy is observed in our (In,Fe)As films [Fig.~\ref{fig:SQUID}(d)] confirming that ferromagnetic nano-lamellae assume the zinc-blende structure for which the $[1{\bar{1}}0]$ and $[110]$ directions are equivalent.

\section{Theoretical description of anisotropic phase separation by {\em ab initio} computations}
\label{sec:ab initio}
\begin{figure}[tb]
\includegraphics[width=8.5 cm]{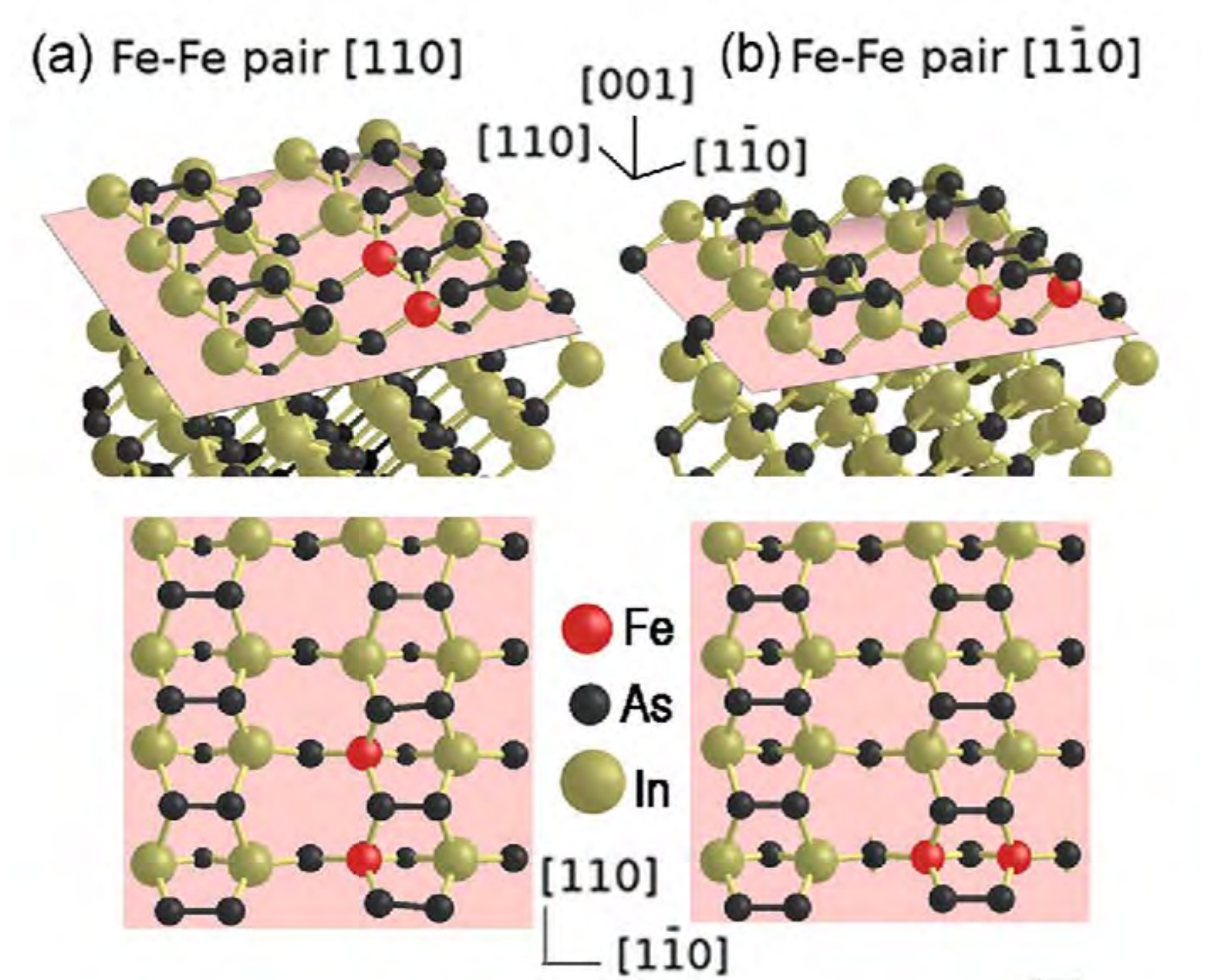}
 \caption{\label{fig:ab initio} (Color online) The optimized structures of different Fe cationic dimer arrangements into (001) InAs surface. The panels (a) and (b) present optimized structures of the Fe pair along the (a) [110] and (b) $[1\bar{1}0]$ crystallographic directions, respectively. The top pictures present the side views of the
slab, whereas the bottom ones are the top views of the surface. The pink plane is presented only for the visibility of the Fe pair in the slab.}%
\end{figure}

We start theoretical interpretation of the observed nematic structure by noting that $[110]$ and  $[1\bar{1}0]$ directions are not equivalent at a (001) surface of zinc-blende compounds and results in the $C_{2v}$ symmetry. In particular, at a cation terminated surface, TM ions are connected by subsurface anions only if they reside along the $[1\bar{1}0]$ direction, as seen comparing Figs.~\ref{fig:ab initio}(a) and \ref{fig:ab initio}(b). This results in a preferential aggregation of TM ions along the $[1\bar{1}0]$ axis, provided that temperature is high enough to make surface diffusion barriers irrelevant \cite{Birowska:2012_PRL}. Within this model, the highly anisotropic Fe distribution, in the form of Fe-rich lamellae, sets-in at the interface between the liquid and recrystallized phase, where Fe cations assume lattice positions minimizing their anion-mediated interaction energies resulting from $p$--$d$ hybridization.

To quantify this model we adopt the previous {\em ab initio} methodology \cite{Birowska:2012_PRL} to the present case. We perform {\em ab initio} calculations employing the SIESTA code \cite{Ordejon:1996_PRB} within the local spin density approximation. A split double zeta basis set with spin polarization (DZP) are used for indium and arsenic atoms, whereas for Fe atoms the triple zeta polarization basis set (TZP) are employed. The kinetic energy cut-off of 200 Ry and $3\times3\times1$ Monkhorst-Pack grid of $k$-points are applied. The size of the supercell is 17\,{\AA} $\times$ 17\,{\AA} $\times$ 43\,{\AA}, consisting of 288 atoms (taking into account pseudo-hydrogen atoms).

Our calculations are divided into two parts. In the first part, we consider (001) InAs surface with the dimerization of the As atoms at the top. The dimerization of the atoms at the surfaces makes the surface more energetically preferable, which is a consequence of reducing the numbers of dangling bonds on the surface by creating $sp^2$ like bonds. To model this surface, slab calculations are performed. The slab consisting of eight double As-In layers (DLs) lay in the (001) crystallographic plane, and 16\,{\AA} of vacuum are used. Each layer contains 16 atoms, making in total 256 atoms in the supercell. The calculated lattice parameter is 6.01\,{\AA} (the experimental one is 6.04\,{\AA}). The dangling bonds from the bottom of the slab are saturated with extra layer consisting of the pseudo-hydrogen atoms with charge equal to $Z = 1.25$ (each indium atom creates two bonds with the pseudo-hydrogen atoms), in order to mimic the bulk types of bonds. All of the positions of the atoms are fully relaxed until the maximal force on each of the atoms reaches a value of 0.02\,eV/{\AA}. In addition, by studying an asymmetric slab one has to deal with a non-vanishing surface-dipole density, due to the fact that the electrostatic potentials at the cell boundaries ((i.e., at the two opposite slab sites) are different. Therefore, the {\em Slab Dipole Correction} flag in the SIESTA code is used in order to get rid of this effect. Moreover, in order to guarantee system neutrality and properly treat the system with a large region of vacuum, the {\em Simulate Doping} flag is also on.

\begin{figure}
\includegraphics[width=8.5 cm]{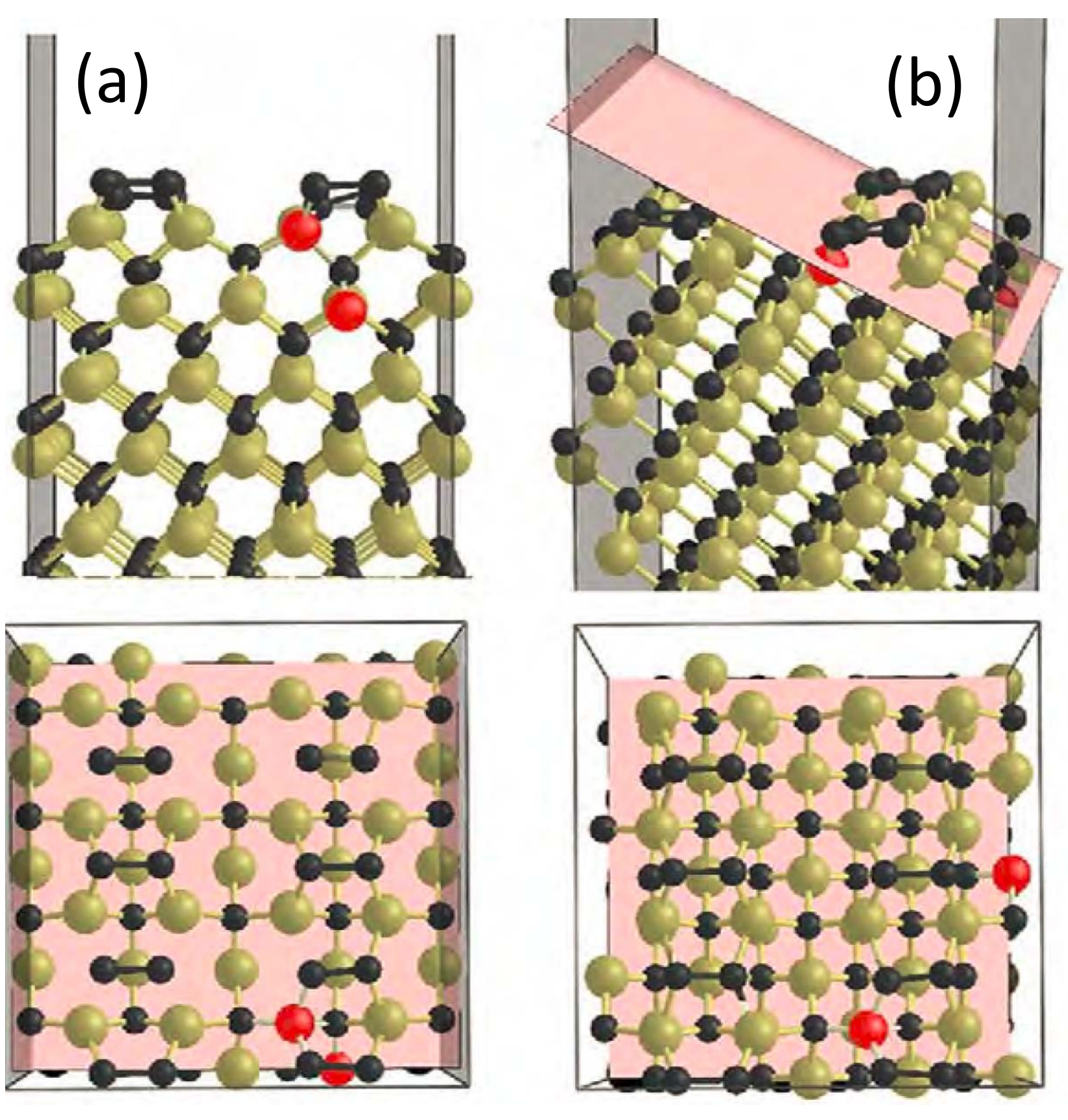}
 \caption{\label{fig:ab_initio2} (Color online) The optimized structures of different Fe cationic dimer arrangements into (001) InAs surface. The (a) and (b) denote positions of Fe cation dimers pointing inside the slab, along the $[01\bar{1}]$ and $[30\bar{1}]$ directions, respectively. The top pictures present the side views of the slab, whereas the bottom ones are the top views of the surface. The pink plane is presented only for the visibility of the Fe pair in the slab. The optimized structures of the Fe pair along the  [110] and  $[1\bar{1}0]$ crystallographic directions are shown in Fig.~\ref{fig:ab initio}.}%
\end{figure}

In the second part of the calculations, we use the optimized reconstructed (001) InAs surface as an input for the (In,Fe)As calculations. We fixed 10 ML from the bottom of the slab, and allowed to relax 6 ML from the top of it. Two indium atoms were substituted by the two Fe atoms at the second top layer of the slab at the nearest-neighbor position. The Fe-coverage of the layer was equal to 1/8 ML.

We find out that the energy of a $[1{\bar{1}}0]$ Fe cation dimer at the (001) InAs surface is lower by 0.11\,eV compared to the [110] Fe cation pair even for the As-terminated surface [Figs.~\ref{fig:ab initio}(a) and \ref{fig:ab initio}(b)]. This result implies that Fe cations tend to aggregate into chains along the $[1{\bar{1}}0]$ direction at the recrystallization front.

Furthermore, as shown in Fig.~\ref{fig:ab_initio2}, we have determined the total energy of the slab containing the nearest Fe cation dimer pointing inside the slab along the $[01\bar{1}]$ direction to the case when the surface Fe cation and the Fe cation in the undersurface plane are at a horizontal distance of 1.6 lattice constants (the largest possible horizontal distance between the Fe cations for the given supercell). We obtain that the energy of the nearest neighbor dimer is lower by 0.35\,eV, which explains why vertical Fe-rich lamellae emerge from the recrystallization process. Furthermore, according to the calculation, the Fe pair exhibits ferromagnetic ordering of the spins in all of the studied configurations (see Figs.~\ref{fig:ab initio} and
\ref{fig:ab_initio2}). It worth mentioning that previous {\em ab initio} computations indicated that the interaction of Fe pairs residing in the bulk is repulsive for the nearest neighbor Fe cations in InAs \cite{Vu:2014_JJAP}.

Therefore, we address the question of whether Fe cation dimers oriented along the $[1\bar{1}0]$ direction are stable at the (001) InAs surface.  Thus, we compute the Fe-Fe pair interaction energy $E_d$ (pairing energy) \cite{Schilfgaarde:2001_PRB,Sato:2005_JJAP,Dietl:2015_RMP,Mahadevan:2005_APL,Gonzalez:2011_PRB},
\begin{equation}
E_d = E_{\text{(In,2Fe)As}}+E_{\text{InAs}} -2E_{\text{(In,Fe)As}},
\end{equation}
where $E$ is the total energy for slabs containing one surface Fe cation dimer, no Fe, and one Fe atom, respectively. The pairing energy is found to be $E_d = -1.16$\,eV per supercell containing 128 cations and 128 anions. Hence, our results show that the surface Fe ions prefer to be close to each other by occupying neighboring In sites at the $[1\bar{1}0]$ crystallographic direction at the (001)InAs surface.

\section{Theoretical description of anisotropic magnetoresistance for $C_{2v}$ symmetry group}
\label{sec:AMR}
The presence of spatially oriented Fe-rich nano-lamellae lowers the film symmetry to the $C_{2v}$ point group, which should be reflected in anisotropy of transport properties. While substantial conductivity of InAs wafers precludes meaningful magnetoresistance studies of our implanted and recrystallized (In,Fe)As layers, such measurements were successfully performed for MBE-grown highly conducting (In,Fe)As:Be films inserted between thin InAs buffer and cap layers deposited onto semiinsulating GaAs substrates \cite{Hai:2012_APL_B}. We show in this section that anisotropic magnetoresistance (AMR) data obtained for epitaxial (In,Fe)As:Be samples are consistent with the $C_{2v}$ symmetry.  This indicates that anisotropic phase separation sets-in during the epitaxial growth of (In,Fe)As:Be.

\subsection{General theory of AMR}

Due to spin-orbit coupling, the magnitude of tensor components describing charge and heat transport depends on the orientation of carrier spin polarization vector ${\cal{P}}$ with respect to symmetry axes and the direction of charge or heat current. In magnetic materials ${\cal{P}}$ is usually collinear with the magnetization vector $\mathbf{M}$ whereas, in non-magnetic cases with the direction of an external magnetic field $\mathbf{H}$. In general, the spin-orbit phenomena compete with effects of the Lorentz force and Landau quantization as well as with spin effects, such as spin disorder scattering present also in the absence of spin-orbit interactions. Of course, the magnitude of all these phenomena is modified by correlation effects, and the implementation of the group theory should consider the existence of spontaneous symmetry breaking in collective phases.

We consider the longitudinal AMR and the planar Hall effect (PHE), i.e.,  the transverse AMR of thin ferromagnetic films as a function of the in-plane directions of magnetization $\mathbf{M}$ and electric current $\mathbf{i}$ \cite{Rushforth:2007_PRL,Birss:1964_B,Rout:2017_PRB} but the formalism is valid also for non-magnetic materials by replacing the direction of $\mathbf{M}$ by the direction of the in-plane component of the magnetic field $\mathbf{H}$.  In this geometry, spin-orbit effects usually account for non-zero AMR and PHE. Both AMR and PHE are described by in-plane resistivity tensor components symmetric in magnetization, $\rho_{ij}(\widehat{\alpha}) = \rho_{ij}(-\widehat{\alpha})$, where the subscripts $i,j \in (1,2) \equiv (x,y)$ correspond to two orthogonal axes of the crystal under considerations, and $\widehat{\alpha} = (\alpha_1,\alpha_2)=(\cos\theta ,\sin\theta )$ is a unit vector in the $\mathbf{M}$ direction in the same reference frame. In general, the in-plane resistivity tensor $\rho_{ij}(\widehat{\alpha})$ can be  expanded into MacLaurin's series,
\begin{equation}
\begin{split}
\label{ro}
\rho_{ij}(\widehat{\alpha} ) =&a_{ij}+a_{klij}\alpha_k \alpha_l+ a_{klmnij}\alpha_k \alpha_l \alpha_m \alpha_n \\
&+a_{klmnpqij}\alpha_k \alpha_l \alpha_m \alpha_n \alpha_p \alpha_q  \\
&+a_{klmnpqrsij}\alpha_k \alpha_l \alpha_m \alpha_n \alpha_p \alpha_q\alpha_r \alpha_s+ .... .
\end{split}
\end{equation}
According to Onsager's relations $\rho_{ij}(\widehat{\alpha}) = \rho_{ji}(\widehat{\alpha})$ in our case and, thus, it is symmetric under any interchange of the indices $k,l,...$ .

The longitudinal resistance, for the in-plane current $\mathbf{i}$  directed along the unit vector $\widehat{\beta} =(\beta_1,\beta_2)=(\cos\varphi ,\sin\varphi)$ in the same reference frame,  is given by,
\begin{equation}
\rho_L\equiv \rho_L(\widehat{\alpha},\widehat{\beta})=\rho_{ij}\beta_i\beta_j=\rho_{11}\beta_1^2+\rho_{22}\beta_2^2+2\rho_{12}\beta_1\beta_2,
\end{equation}
whereas the planar Hall effect by,
\begin{equation}
\begin{split}
\rho_{PH}&\equiv\rho_{PH}(\widehat{\alpha},\widehat{\beta})=\rho_{ij}\beta'_i\beta_j \\ &=\rho_{11}\beta_1'\beta_1+\rho_{22}\beta_2'\beta_2+\rho_{12}\beta_1'\beta_2+\rho_{21}\beta_2'\beta_1,
\end{split}
\end{equation}
where $\widehat{\beta'}=(-\sin\varphi,\cos\varphi)$ is a unit vector perpendicular to $\mathbf{i}$.

\begin{figure}[h]
		\includegraphics[width=5cm]{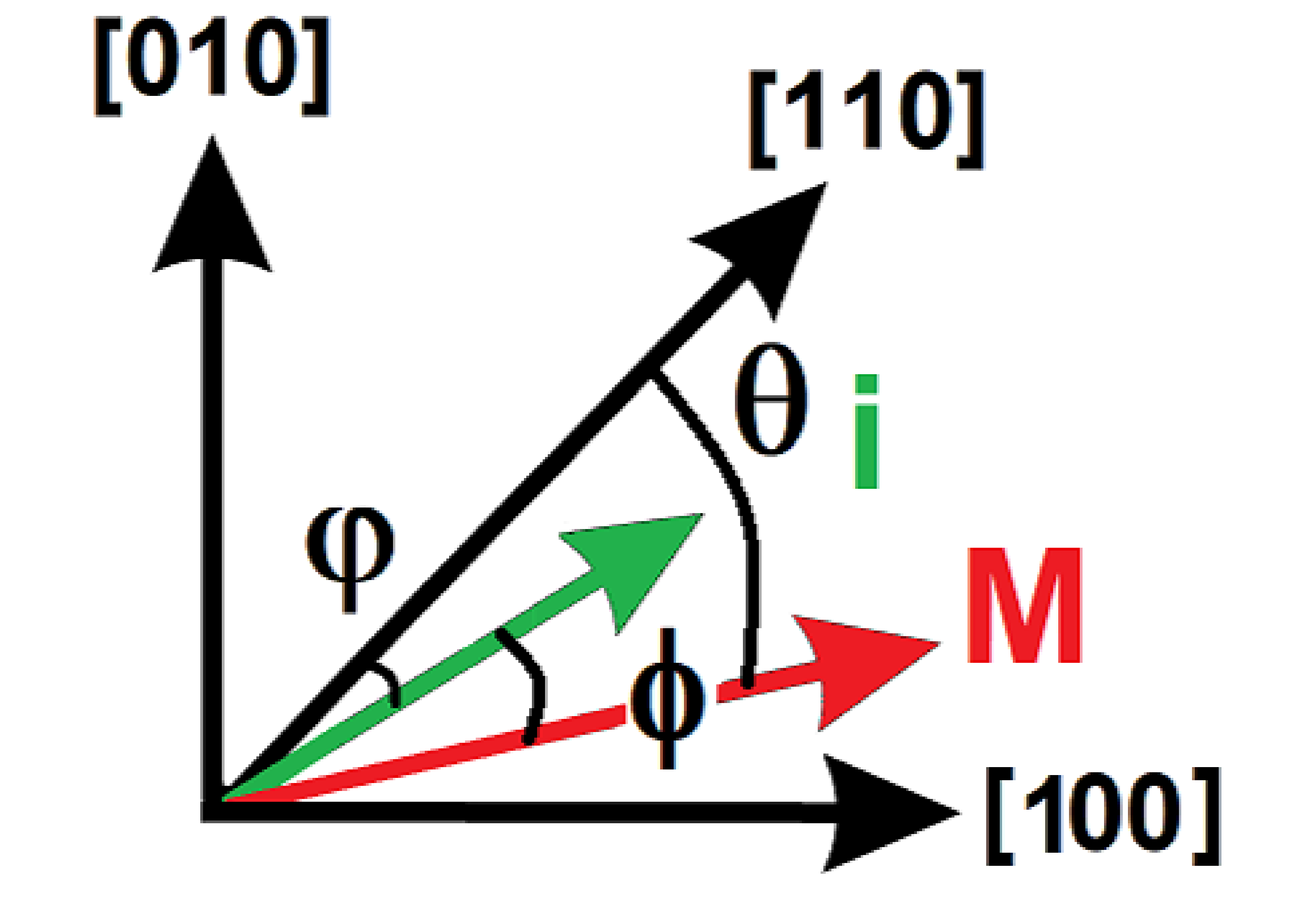}
	\caption{\label{fig:ax} Definitions of the angles.}
\end{figure}

We derive formulas for the longitudinal and transverse resistances (up to the eighth order) for the case of thin films possessing the $C_{2v}$ point group symmetry, as implied by the presence of Fe-rich lamellae oriented along the $[1{\bar{1}}0]$ in-plane direction. As detailed in Appendix, the longitudinal resistance assumes the form,
\begin{equation}
\label{rhoL}
\begin{split}
\rho_L(\theta, \varphi)=&\rho_{xx}(\theta)\cos^2(\varphi)+\rho_{yy}(\theta)\sin^2(\varphi)\\&
+2\rho_{xy}(\theta)\cos(\varphi)\sin(\varphi),
\end{split}
\end{equation}
where the angles ($\theta$, $\varphi$) are defined in reference to the [110] crystallographic direction (see Fig.~\ref{fig:ax}).

The planar Hall effect can be then expressed as,
\begin{equation}
\label{rhoH}
\begin{split}
\rho_{PH}(\theta, \varphi)=&-\frac{1}{2}\rho_{xx}(\theta)\sin(2\varphi)+\frac{1}{2}\rho_{yy}(\theta)\sin(2\varphi)\\&
+\rho_{xy}(\theta)\cos(2\varphi).
\end{split}
\end{equation}

Now we introduce angle $\phi$ between the directions of the magnetization $\mathbf{M}$ and the current $\mathbf{i}$. The  configuration involving $\theta$ and $\phi$ angles is commonly used in experimental setups. The coordinate system and the definitions of particular angles are presented in Fig.~\ref{fig:ax}. Hence, the longitudinal and transverse resistivity assumes the form,
  \begin{equation}
 \label{rhoLN}
 \begin{split}
 \rho_L(\theta, \phi)=&\rho_{xx}(\theta)\cos^2(\theta-\phi)+\rho_{yy}(\theta)\sin^2(\theta-\phi)\\ &
 +\rho_{xy}(\theta)\sin[2(\theta-\phi)],
 \end{split}
 \end{equation}
 \begin{equation}
 \begin{split}
 \label{rhoHN}
 &\rho_{PH}(\theta, \phi)=-\frac{1}{2}\rho_{xx}(\theta)\sin[2(\theta-\phi)]\\
 &+\frac{1}{2}\rho_{yy}(\theta)\sin[2(\theta-\phi)] +\rho_{xy}(\theta)\cos[2(\theta-\phi)].
 \end{split}
 \end{equation}
 The dependence of $\rho_L$ and $\rho_{PH}$ on $\theta$ at given $\phi$ corresponds to the "crystalline" AMR, whereas on $\phi$ at given $\theta$ to the  "noncrystalline" AMR.

We consider two cases, when the current $\mathbf{i}$ is parallel to $[110]$ and $[\bar{1}10]$ axes, respectively. Furthermore, we compare our findings to experimental results reported in Ref.~\onlinecite{Hai:2012_APL_B} for (In,Fe)As:Be films grown by MBE on semi-insulating GaAs substrates.

\subsection{AMR -- current \texorpdfstring{$\mathbf{i}\parallel [110]$ direction}{TEXT}}

\begin{figure}
	\includegraphics[width=8cm]{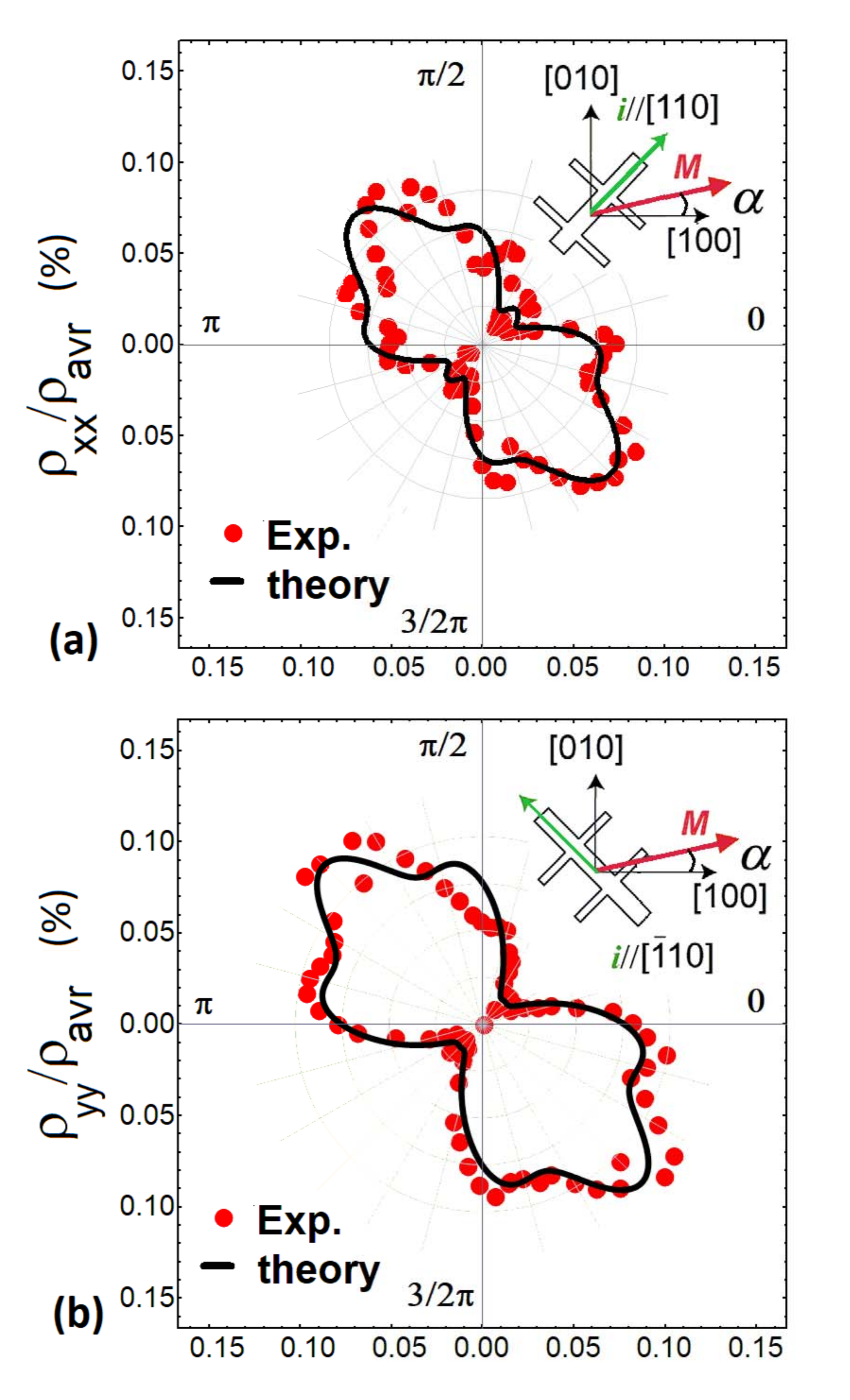}
	\caption{\label{fig:AMR} Polar plots of the longitudinal anisotropic magnetoresistance as a function of the angle $\alpha$ defined in the insets: (a) current parallel to [110]  and (b) current parallel to $[\bar{1}10]$ crystallographic directions. Red dots are experimental data for (In,Fe)As:Be taken from Ref.~\cite{Hai:2012_APL_B}. Solid black lines are fitted curves with the parameters given in the text.}
\end{figure}

When current is parallel to the [110] axis then $\phi=\theta$ (see Fig.~\ref{fig:ax}). One can easily see that the second and the third terms in the Eq.~\ref{rhoLN} vanish, and hence, the longitudinal magnetoresistance has the form,
\begin{equation}
\begin{split}
\rho_L(\theta, \phi=\theta)\equiv\rho_{xx}=&C_0 +C_2\cos(2\theta)+C_4\cos(4\theta) \\
&+C_6\cos(6\theta)+C_8\cos(8\theta),
\end{split}
\end{equation}
which we use to describe experimental results reported in Ref.~\cite{Hai:2012_APL_B} for (In,Fe)As:Be grown by MBE.
As shown in Fig.~\ref{fig:AMR}(a), we have fitted our theory to 69 data points.  Since the model considered previously \cite{Hai:2012_APL_B} considered only $C_2, C_4, C_8$ terms and estimated the $C_i$  values from peak positions, the coefficients determined  here, $C_0=0.055\%, C_2=-0.034\%, C_4=-0.0007\%, C_6=-0.002\%, C_8= 0.008\% $, are more realistic. The two largest correspond to the two- and eight-fold contributions, in agreement with the previous conclusion \cite{Hai:2012_APL_B}. A non-zero value of $C_2$ points to the $C_{2v}$ point symmetry, whereas relative magnitudes of other coefficients reflect details of the nano-lamella arrangement, and vary from sample to sample. In particular, in the case of samples with a higher electron concentration, the two-fold character of the AMR dominates \cite{Hai:2012_APL_B}, i.e., $|C_2| \gg |C_i|$, where $i = 4, 6$, and 8.

\subsection{AMR -- current \texorpdfstring{$\mathbf{i}\parallel [\bar{1}10]$ direction}{TEXT}}
When current is parallel to $[\bar{1}10]$ then $\phi=\theta+\pi/2$, and hence, the longitudinal magnetoresistance has the form
\begin{equation}
\begin{split}
\rho_L(\theta, \phi=\theta+\pi/2)\equiv\rho_{yy}=&D_0+D_2\cos(2\theta)+D_4\cos(4\theta) \\
&+D_6\cos(6\theta)+D_8\cos(8\theta).
\end{split}
\end{equation}
In this case we have fitted our model to 70 data points reported in Ref.~\cite{Hai:2012_APL_B}. The coefficients obtained from the best fit are $D_0=0.066\%, D_2=-0.047\%, D_4=-0.005\%, D_6=-0.002\%, D_8=0.008\%$. The fitted curve is presented in the Fig.~\ref{fig:AMR}(b).
We want to notice  that although the longitudinal magnetoresistance for current along either $[110]$ or $[\bar{1}10]$ has the same cosine terms, the coefficients are not equal. This may be assigned to an expected inequality between $\sigma_{xx}$ and $\sigma_{yy}$ in the $C_{2v}$ case and/or may originate from differences in the details of Fe distributions in these two devices.

\subsection{Planar Hall effect }
\begin{figure}[h!]
	\includegraphics[width=8cm]{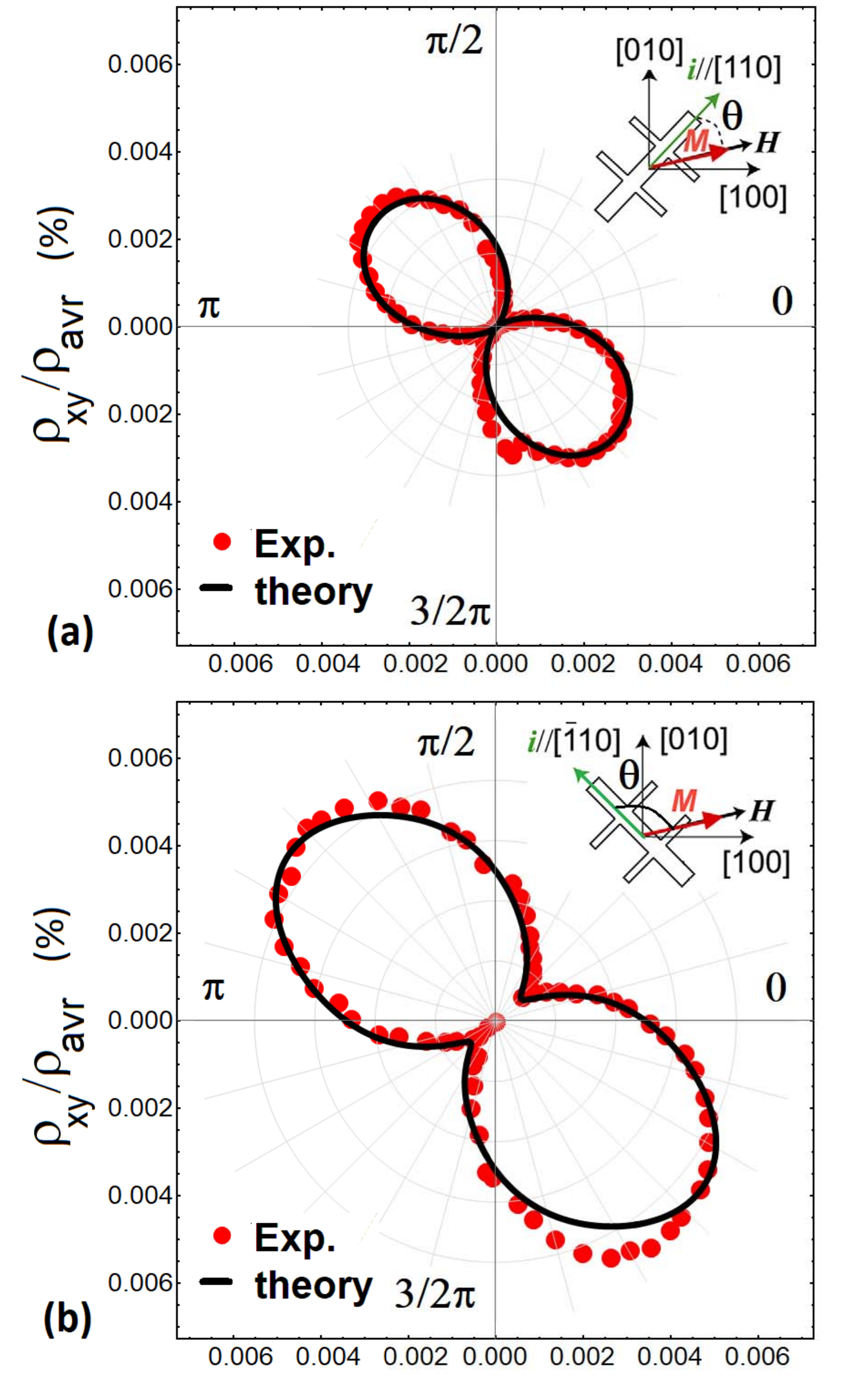}
	\caption{Polar plots of the planar Hall effect as a function of the angle $\theta$ (defined in the insets) for: (a) current parallel to [110]  and (b) current parallel to $[\bar{1}10]$ crystallographic directions. Red dots are experimental data for (In,Fe)As:Be taken from Ref.~\cite{Hai:2012_APL_B}, whereas solid black lines are fitted curves with the parameters given in the text.}
\label{fig:PHE}
\end{figure}

We also present the results for the planar Hall effect for the same two current orientations: when the current is parallel to [110] crystallographic direction then $\rho_{PH}(\phi=\theta)=\rho_{xy}$, and if the current is directed along $[\bar{1}10]$  then $\rho_{PH}(\phi=\theta+\pi/2)=-\rho_{xy}$ (see Eq.~\ref{rhoHN}).
The fitted curves are presented in Fig.~\ref{fig:PHE} as a function of $\theta$ for samples showing primarily the two-fold character in the longitudinal AMR \cite{Hai:2012_APL_B}. The parameters obtained from the best fits for $\mathbf{i} \parallel [110]$, and $\mathbf{i} \parallel [\bar{1}10]$ are $S_2=-1.8\times10^{-3}$\%, $S_4=S_6=S_8=0$, and  $S_2=2.5\times10^{-3}$\%, $S_4=0.3 \times10^{-3}$\%, $S_6=S_8=0$, respectively. In both cases an angle-independent parameter has been added to the fits, which does not result from the symmetry consideration, as according to the formula for $\rho_{xy}$ in Eq.~\ref{terms}, there is no such a term. It may arise from an experimental off-set. Furthermore, while---as expected---the signs of $\rho_{PH}(\phi=\theta)$ and $\rho_{PH}(\phi=\theta+\pi/2)$ are reversed, the absolute values of the PHE signals are not equal indicating the existence of certain differences in details of Fe distributions in the two studied devices.

\section{Discussion of nematicity origin in $\mbox{(In,Fe)As, (Ga,Mn)As}$, and other systems}
\subsection{Comparison of recrystallized $\mbox{(In,Fe)As}$ and epitaxial $\mbox{(In,Fe)As:Be}$}
As could be expected, no evidences for Fe aggregation was found for (In,Fe)As samples obtained by LT-MBE \cite{Hai:2012_APL_A}, as under such growth conditions slow surface diffusion precludes the thermal equilibrium distribution of Fe cations. Similarly to Mn$^{2+}$ in II-VI compounds, the Fe$^{3+}$ ions act in InAs as substitutional randomly distributed isoelectronic impurities that, due to large inter-spin distances,  are weakly coupled by exchange interactions \cite{Hai:2012_APL_A} that are expected to be ruled by short-range antiferromagnetic superexchange in the $d^5$ case \cite{Dietl:2014_RMP}.

We claim, however, that the distribution of Fe cations in Be-doped (In,Fe)As epitaxial films \cite{Hai:2012_APL_A} is similar to that found in our recrystallized (In,Fe)As samples, i.e., it assumes the form of spatially oriented Fe-rich nano-lamellae presented in Figs.~\ref{fig:TEM} and \ref{fig:EDXS}, whose formation mechanism is explained with the aid of {\em ab initio} computations in Sec.~\ref{sec:ab initio}.  Indeed, these two systems show similar magnitudes of blocking temperature and the behavior of thermoremanent magnetization \cite{Hai:2012_APL_A}. Most importantly, as already shown in Sec.~\ref{sec:AMR}, we can explain, by extending to eighth order the AMR theory for the $C_{2v}$ crystal symmetry, the surprising nematicity of (In,Fe)As:Be, as revealed by two-fold and eight-fold "crystalline" AMR \cite{Hai:2012_APL_B}. The $C_{2v}$ symmetry is actually experienced by current-carrying electrons moving in-plane of a (001) zinc-blende film with lamellae parallel to the (110) surface.  Good agreement between the experimental and theoretical data on the longitudinal and transverse AMR, i.e., the planar Hall effect (PHE), together with the magnitudes of the determined parameters, demonstrates a strong symmetry breaking, $D_{2d} \rightarrow C_{2v}$, in MBE-grown (In,Fe)As:Be films \cite{Hai:2012_APL_B}. This points to the presence of unidirectionally oriented lamellae also in those films and explains why "crystalline" terms dominate the character of AMR.

It might appear that the presence of spatially oriented ferromagnetic Fe-rich nano-lamellae should lead to strong and nematic crystalline magnetic anisotropy. Actually this is {\em not} the case, as ferromagnetism originates from {\em cubic} zinc-blende Fe-rich nano-regions. Since in the cubic case the [110] and $[\bar{1}10]$ crystal axes are equivalent, no differences in the behavior of $M(H)$ are expected for the magnetic field $\bm{H}$ oriented in either of these two in-plane directions, as observed [Fig.~\ref{fig:SQUID}(d)]. For the same reason, there is no crystalline magnetic anisotropy between [100] and [001] crystal axes, so that in this case the shape anisotropy dominates in the (001) films, as indeed found for epitaxial (In,Fe)As:Be layers \cite{Hai:2012_APL_A}. As shape magnetic anisotropy is driven by long range dipole interactions between magnetic moments, its magnitude is determined by the value of average magnetization and by specimen shape, whereas a non-random distribution of Fe ions at the nanoscale is irrelevant in this case.

\subsection{$\mbox{(In,Fe)As}$ vs. $\mbox{(Ga,Mn)As}$}

Experimental results accumulated for recrystallized and epitaxial films of (In,Fe)As and (In,Fe)As:Be, respectively demonstrate that these systems constitute a markedly different class of materials compered to (Ga,Mn)As and (Ga,Mn)As:Be obtained by the same methods.  In particular, the previous and present studies reveal five major differences.

First, (Ga,Mn)As and (Ga,Mn)As:Be are $p$-type, in which valence band holes  mediate ferromagnetic coupling between spins localized on randomly distributed Mn ions \cite{Dietl:2014_RMP,Jungwirth:2014_RMP}. This is in contrast to (In,Fe)As and (In,Fe)As:Be, which are $n$-type \cite{Hai:2012_APL_A} but nevertheless recrystallized (In,Fe)As layers (Fig.~\ref{fig:SQUID}) and MBE-grown (In,Fe)As:Be films \cite{Hai:2012_APL_A} show ferromagnetic features. Second, the  "noncrystalline" AMR takes over in (Ga,Mn)As \cite{Rushforth:2007_PRL},  in contrast to the case of (In,Fe)As:Be in which symmetry breaking two-fold crystalline AMR prevails \cite{Hai:2012_APL_B} and points to nematic $C_{2v}$ symmetry, as shown in Figs.~\ref{fig:AMR} and \ref{fig:PHE}. Third, magnetic anisotropy of (Ga,Mn)As and (Ga,Mn)As:Be is governed by crystalline contributions that break tetragonal symmetry of the films \cite{Dietl:2014_RMP}, whereas no breaking of cubic magnetic anisotropy, i.e., {\em no} nemacity of magnetic anisotropy, is found in recrystallized (In,Fe)As [Fig.~\ref{fig:SQUID}(d)] and (In,Fe)As:Be \cite{Hai:2012_APL_A}.  Fourth, in contrast to the case of Mn in (Ga,Mn)As and (Ga,Mn)As:Be, the distribution of Fe ions is highly nonuniform in recrystallized films of (In,Fe)As (Figs.~\ref{fig:TEM} and \ref{fig:EDXS}) and in MBE-grown (In,Fe)As:Be \cite{Sakamoto:2016_PRB}. Finally, a substantial concentration of interstitial Mn donors was detected in (Ga,Mn)As \cite{Yu:2002_PRB} and (Ga,Mn)As:Be \cite{Wojtowicz:2003_APLb}, whereas no Fe interstitials are found by our Rutherford backscattering studies of recrystallized (In,Fe)As (Sec.~\ref{sec:SIMS}).

We argue that the above differences result from the opposite sign of the charge transfer energy $E_{\text{ct}}$ in these two classes of magnetic semiconductors. Because of the intra-ion exchange coupling among $d$ electrons, this energy, i.e., the position of the TM acceptor state with respect to the top of the valence band, is negative for Mn in GaAs but positive for Fe in InAs \cite{Dietl:2002_SST}. This accounts for the acceptor nature of Mn in GaAs, and the isoelectronic character of Fe in InAs. However, compared to widely studied II-VI compounds with Mn in the $d^5$ configuration,  the magnitude of $E_{\text{ct}}$ is much smaller in the Fe case, so that the Fe acceptor level resides in the bandgap or only slightly higher than the bottom of the conduction band in InAs \cite{Dietl:2002_SST}. By this fact, we explain why Be acts as a substitutional acceptor in InAs but as an interstitial donor in (In,Fe)As, as established for films grown by MBE \cite{Hai:2012_APL_A}. Indeed, trapping of electrons by Fe allows us to avoid an increase of the system energy associated with either reducing the number of electrons in the bonding states (i.e., introducing holes to the valence band) or increasing the number of electrons in the antibonding states (i.e., occupying the conduction band). Experimental results imply also that the process of interstitial formation, accompanied by the appearance of cation vacancies and charging of Fe ions, facilitates the aggregation of Fe even under MBE growth conditions, i.e., at relatively low temperatures. The assembling of Fe-rich nano-lamellae is associated with a release of electrons provided by Be to the conduction band, making the material $n$-type \cite{Hai:2012_APL_A}.

The aforementioned formation of interstitial donors occurs also in the case of (Ga,Mn)As and (Ga,Mn)As:Be but involves primarily Mn acceptors \cite{Yu:2002_PRB,Wojtowicz:2003_APLb}. This process reduces the hole concentration and $T_{\text{C}}$ of (Ga,Mn)As \cite{Yu:2002_PRB} and also of (Ga,Mn)As:Be provided that co-doping by Be is performed during the epitaxy of (Ga,Mn)As \cite{Wojtowicz:2003_APLb}. The preferential aggregation of Mn along  the $[1\bar{1}0]$ direction leads to in-plane uniaxial crystalline magnetic anisotropy \cite{Birowska:2012_PRL}, essential for many celebrated spintronic functionalities of (Ga,Mn)As \cite{Dietl:2014_RMP,Jungwirth:2014_RMP}. However, the magnetic anisotropy magnitude implies a low degree of anisotropic phase separation \cite{Birowska:2012_PRL} and, accordingly,  it can hardly be visualized directly by nanocharacterization tools, and affects rather weakly AMR of (Ga,Mn)As \cite{Rushforth:2007_PRL}.

\subsection{Effects of chemical phase separation in other systems}
Having elucidating the origin of nematicity, i.e., of the rotational symmetry breaking $D_{2d} \rightarrow C_{2v}$ in (In,Fe)As and (Ga,Mn)As we address the question of to what extent chemical phase separation might account for the nematic characteristic of other systems. In particular, we consider the case of unconventional superconductors and III-V semiconductors in the quantum Hall effect regime.

Interestingly, lamella-like structures of Fe-rich K$_{x}$Fe$_{2}$Se$_2$ extending along $[1{\bar{1}}0]$  and $[110]$ directions, and surrounded be an Fe-poor K$_{x}$Fe$_{2-y}$Se$_2$ matrix with a different crystal structure were identified by nanocharacterization tools \cite{Wang:2012_JPCC,Tanaka:2017_APEX,Pomjakushina:2017_B}. These lamellae are formed by chemical phase separation \cite{Pomjakushina:2017_B} and account for the superconducting phase transition with an onset at temperature as high as 44\,K \cite{Tanaka:2017_APEX}.

It was also shown that interactions between impurities intercalated into the van der Waals gaps of Bi$_2$Se$_3$ result in the formation of impurity stripe domains \cite{Mann:2014_PRB,Wang:2016_JPCM}. It is then natural to suggest that these stripes account for the orientation of the nematic axis  in the superconducting phase of Cu-, Nb,-, and Sr-doped topological insulator Bi$_2$Se$_3$ (Refs.~\onlinecite{Matano:2016_NP,Yonezawa:2017_NP,Asaba:2017_PRX,Pan:2016_SR,Du:2017_SCPMA}), strong candidate materials for the unmatched class of odd parity superconductors \cite{Fu:PRL_2010}.

It can be expected that anisotropic chemical phase separation revealed here for (In,Fe)As is an immanent property of epitaxial growth of zinc-blende alloys, though its experimental relevance will, of course, depend on alloy components and growth parameters. We note that a highly anisotropic ripple structure was found in GaAs with vacancies, and assigned to the anisotropy of surface diffusion along $[1{\bar{1}}0]$ and [110] directions \cite{Ou:2015_NS}, the mechanism that may contribute also to the anisotropic phase separation reported here. Epitaxy-induced anisotropy in the angular distribution of alloy components, impurities or defects---acting often together with structural inversion asymmetry---lowers rotational symmetry of epilayers and quantum wells to $C_{2v}$. This leads to apparent nematicity that can manifest itself as, for instance, a difference in the magnitude of  in-plane longitudinal resistance measured along the [110] and $[1{\bar{1}}0]$ directions, $\rho_{xx} \ne \rho_{yy}$. An interesting question arises on whether this quenched anisotropy, or rather spontaneous symmetry breaking by a charge density formation, accounts for $\rho_{xx} \ne \rho_{xx}$ revealed in Ga(Al)As/Al$_x$Ga$_{1-x}$As quantum wells under quantum Hall effect conditions \cite{Fradkin:2010_ARCMP,Hossain:2018_PRB}. Of course, in addition to the "crystalline" effect, a "non-crystalline" AMR is brought about by an in-plane component of the magnetic field. As the latter originates from  spin-orbit interactions, its magnitude is expected to be highly sensitive to the degree of carrier spin-polarization, as observed \cite{Hossain:2018_PRB}.

\section{Summary and outlook}	

Our results show that particular patterns in the quenched distribution of alloy components or dopants hosting correlated carrier or spin systems may either stabilize a spatially non-uniform collective phase with high critical temperature (e.g., In$_{1-x}$Fe$_x$As and K$_x$Fe$_{2-y}$Se$_2$) or account only for the presence and orientation of the two-fold easy axis in a collective phase that exists independently of phase separation (e.g., Ga$_{1-x}$Mn$_x$As and probably Cu$_x$Bi$_{2}$Se$_3$ and Ga(Al)As/Al$_x$Ga$_{1-x}$As quantum wells). From another perspective, these findings, together with previous demonstrations of self-organized assembly of periodically distributed TM-rich nanocolumns embedded in the TM-poor host, the case of epitaxial Ge$_{1-x}$Mn$_x$ \cite{Jamet:2006_NM,Bougeard:2009_NL} and Zn$_{1-x}$Cr$_x$Te films \cite{Nishio:2009_MRSP}, open prospects for self-organized fabrications of functional hybrid structures consisting of ordered metallic, magnetic or superconductive nanostructures embedded in various hosts.

\section*{Acknowledgments}

Supports by the Ion Beam Center (IBC) at HZDR and the funding of TEM Talos by the German Federal Ministry of Education of Research (BMBF), Grant No.~03SF0451 in the framework of HEMCP are gratefully acknowledged. This work is funded by the Helmholtz-Gemeinschaft Deutscher Forschungszentren (HGF-VH-NG-713). Y.\,Y. gratefully
acknowledges the financial support by Chinese Scholarship Council (File No.~201306120027). The work in Poland is supported by the National Science Centre through projects MAESTRO (2011/02/A/ST3/00125) and by the Foundation for Polish Science through the IRA Programme financed by EU within SG OP Programme. M.\,B. is funded by the National Science Centre Grant No.~UMO-2016/23/D/ST3/03446. Financial support by the EU 7th Framework Programme under the project REGPOT-CT-2013-316014 (EAgLE) and the international project co-financed by the Polish Ministry of Science and Higher Education, Grant Agreement No. 2819/7.PR/2013/2 are acknowledged. Access to computing facilities of PL-Grid Polish Infrastructure for Supporting Computational Science in the European Research Space and of the Interdisciplinary Center of Modeling (ICM), University of Warsaw is gratefully acknowledged.

\appendix
\section{Derivation of AMR formulae for $C_{2v}$ symmetry group}
According to the Neumann's principle, the coefficient tensors $a_{ij}, a_{klij}, ...$ in Eq.\,(\ref{ro}) should be unchanged under symmetry operations using the generating matrix $U$,
\begin{equation}
a_{ijkl...}=U_{ip}U_{jq}U_{kr}U_{ls}...a_{pqrs...}.
\end{equation}
The $C_{2v}$ point group contains symmetry operations $U=\left \{E,C_2,\sigma_{xz},\sigma_{yz}\right \}$, where $E$ is the identity matrix  and
\begin{equation}
 C_2=\begin{pmatrix}
-1 &0 \\
0 & -1
\end{pmatrix};
\sigma_{xz}=\begin{pmatrix}
1 &0 \\
0 & -1
\end{pmatrix};
\sigma_{yz}=\begin{pmatrix}
-1 &0 \\
0 & 1
\end{pmatrix}.
\end{equation}

The non-vanishing coefficients in Eq.~\ref{ro} and the relations between them are given in Tables I-V. They have been determined by considering symmetry properties of $\rho_{ij}$. In particular, due to symmetries described by $\sigma_{xz},\sigma_{yz}$ only coefficients with an even number of indices $i,j,k,l,...$ equal to 1 and even number equal to 2 do not vanish.

Accordingly, the resistivity tensor in question can be expressed as
\begin{equation}
\label{terms}
\begin{split}
&\text{\footnotesize $\rho_{xx}=C_0+C_2\cos(2\theta)+C_4\cos(4\theta)+C_6\cos(6\theta)+C_8\cos(8\theta)$}; \\
&\text{\footnotesize $\rho_{yy}=D_0+D_2\cos(2\theta)+D_4\cos(4\theta)+D_6\cos(6\theta)+D_8\cos(8\theta)$}; \\
&\text{\footnotesize $\rho_{xy}=S_2\sin(2\theta)+S_4\sin(4\theta)+S_6\sin(6\theta)+S_8\sin(8\theta)$},
\end{split}
\end{equation}
where the coefficients $C_0, C_2, ...,D_0, D_2, ...,S_2, S_4$ are given by,

\begin{widetext}
\begin{equation}
\begin{split}
C_0 = &\frac{1}{128}(128 a_{11} + 64 a_{1111} +48 a_{111111}+ 35 a_{1111111111} + 140 a_{1111112211}+ 210 a_{1111222211} + 96 a_{112211} + 140 a_{1122222211}  \\
&+ 64 a_{2211} + 48 a_{222211} + 35 a_{2222222211}) \\
C_2 = &\frac{1}{16} (8 a_{1111} + 8 a_{111111} + 7 a_{1111111111} + 14 a_{1111112211}- 14 a_{1122222211} - 8 a_{2211} - 8 a_{222211} - 7 a_{2222222211}) \\
C_4 =& \frac{1}{32} (4 a_{111111} + 7 a_{1111111111} - 28 a_{1111112211} - 70 a_{1111222211} - 24 a_{112211} - 28 a_{1122222211} + 4 a_{222211} + 7 a_{2222222211}) \\
C_6 =& \frac{1}{16} (a_{1111111111} - 14 a_{1111112211} + 14 a_{1122222211} - a_{2222222211}) \\
C_8 =&\frac{1}{128} (a_{1111111111} - 28 a_{1111112211} + 70 a_{1111222211} - 28 a_{1122222211} + a_{2222222211}) \\
D_0=&\frac{1}{128} (35 a_{1111111122} + 2 (20 a_{11111122} + 70 a_{1111112222} + 24 a_{111122} + 60 a_{11112222} + 105 a_{1111222222}+ 32 a_{1122}\\
& + 48 a_{112222} + 60 a_{11222222} + 70 a_{1122222222} + 64 a_{22}+ 32 a_{2222} + 24 a_{222222} + 20 a_{22222222}) + 35 a_{2222222222}) \\
D_2=&\frac{1}{32} (14 a_{1111111122} + 15 a_{11111122} + 28 a_{1111112222} + 16 a_{111122} + 15 a_{11112222} + 16 a_{1122}\\
&- 15 a_{11222222} - 28 a_{1122222222} - 16 a_{2222} - 16 a_{222222} - 15 a_{22222222} - 14 a_{2222222222})\\
D_4=&\frac{1}{32} (7 a_{1111111122} + 6 a_{11111122} - 28 a_{1111112222} + 4 a_{111122} -30 a_{11112222}- 70 a_{1111222222} - 24 a_{112222} - 30 a_{11222222}\\
&-28 a_{1122222222} + 4 a_{222222} + 6 a_{22222222} + 7 a_{2222222222})\\
D_6=&\frac{1}{32} (2 a_{1111111122} + a_{11111122} - 28 a_{1111112222} - 15 a_{11112222} +15 a_{11222222} + 28 a_{1122222222} - a_{22222222} - 2 a_{2222222222}) \\
D_8=&\frac{1}{128}  (a_{1111111122} - 28 a_{1111112222} + 70 a_{1111222222} - 28 a_{1122222222} + a_{2222222222}) \\
S_2=&\frac{1}{16} (14 a_{1111111212} + 15 a_{11111212} + 42 a_{1111122212} + 16 a_{111212} + 30 a_{11122212} + 42 a_{1112222212} + 16 a_{1212} + 16 a_{122212}\\
&+ 15 a_{12222212} + 14 a_{1222222212}) \\
S_4=&\frac{1}{8} (7 a_{1111111212} + 6 a_{11111212} + 7 a_{1111122212} + 4 a_{111212} - 7 a_{1112222212} - 4 a_{122212} - 6 a_{12222212} - 7 a_{1222222212}) \\
S_6=&\frac{1}{16} (6 a_{1111111212} + 3 a_{11111212} - 14 a_{1111122212} - 10 a_{11122212} - 14 a_{1112222212} + 3 a_{12222212} + 6 a_{1222222212}) \\
S_8=&\frac{1}{16} (a_{1111111212} - 7 a_{1111122212} + 7 a_{1112222212} - a_{1222222212}) \\
	\end{split}
	\end{equation}
\end{widetext}

\begin{table}[h!]
		\label{T0}
	\caption{$a_{klij}$ coefficients of Eq.~\ref{ro} ($0\theta$ order)}
\begin{tabular}{|c|ccc|}\hline
		\hline
	\thead{$\beta_i \beta_j$}&\thead{$\beta_1^2$}&\thead{$\beta_2^2$}& 	\thead{$\beta_1\beta_2$} (=$\beta_2\beta_1$)\\
		\hline
		$ij$& $a_{11}$ & $a_{22}$ &  \\
    \hline
	\end{tabular}
\end{table}

\begin{table}[h!]
	\label{T2}
	\caption{$a_{klij}$ coefficients of Eq.~\ref{ro} ($2\theta$ order).}
\begin{tabular}{|c|ccc|}\hline
	\diaghead{\theadfont Diag ColumnmnHead II}%
		{$\alpha_k \alpha_l$}{\\$\beta_i\beta_j$}&\thead{$\beta_1^2$}&\thead{$\beta_2^2$}& 	\thead{$\beta_1\beta_2$} (=$\beta_2\beta_1$)\\
		\hline
			& & &\\
		$\alpha_1^2$& $a_{1111}$ & $a_{1122}$ &  \\
		$\alpha_2^2$& $a_{2211}$ & $a_{2222}$ &  \\
		$\alpha_1\alpha_2$ (x2)&  &  &$a_{1212}$ \\
	
		\hline
		\end{tabular}
		\end{table}

\begin{table}[h!]
	\label{T4}
	\caption{$a_{klmnij}$ coefficients of Eq.~\ref{ro} ($4\theta$ order)}
	\begin{tabular}{|c|ccc|}\hline
		\diaghead{\theadfont Diag ColumnmnHead II}%
		{$\alpha_k \alpha_l\alpha_m \alpha_n$}{\\$\beta_i\beta_j$}&\thead{$\beta_1^2$}&\thead{$\beta_2^2$}& 	\thead{$\beta_1\beta_2$}  (=$\beta_2\beta_1$)\\
		\hline
			& & &\\
		$\alpha_1^4$& $a_{111111}$ & $a_{111122}$ &  \\
		$\alpha_2^4$ & $a_{222211}$ & $a_{222222}$ & \\
		$\alpha_1^2\alpha_2^2$ (x6)& $a_{112211}$ &$a_{112222}$  &   \\
		$\alpha_1^3\alpha_2$ (x4)&  &  & $a_{111212}$ \\
		$\alpha_1\alpha_2^3$ (x4)&  &  &$a_{122212}$ \\
				\hline
	\end{tabular}
\end{table}

\begin{table}[h!]
	\label{T6}
	\caption{$a_{klmnij}$ coefficients of Eq.~\ref{ro} ($6\theta$ order)}
	\begin{tabular}{|c|ccc|}\hline
		\diaghead{\theadfont Diag ColumnmnHead II}%
		{$\alpha_k \alpha_l\alpha_m \alpha_n\alpha_p\alpha_q$}{\\$\beta_i\beta_j$}&\thead{$\beta_1^2$}&\thead{$\beta_2^2$}& 	 \thead{$\beta_1\beta_2$}   (=$\beta_2\beta_1$)\\
		\hline
		& & &\\
		$\alpha_1^6$& $a_{11111111}$ & $a_{11111122}$ &  \\
		$\alpha_2^6$ & $a_{22222211}$ & $a_{22222222}$ &  \\
		$\alpha_1^4\alpha_2^2$ (x15)&  $a_{11112211}$&$a_{11112222}$  &  \\
		$\alpha_1^2\alpha_2^4$ (x15)& $a_{11222211}$ &$a_{11222222}$  &   \\
		$\alpha_1^5\alpha_2$ (x6)&  &  &$a_{11111212}$ \\
		$\alpha_1^3\alpha_2^3$ (x20)&  &  &$a_{11122212}$ \\
		$\alpha_1\alpha_2^5$ (x6)&  &  &$a_{12222212}$ \\
			\hline
	\end{tabular}
\end{table}

\begin{table}[h!]
	\label{T8}
	\caption{$a_{klmnij}$ coefficients of Eq.~\ref{ro} ($8\theta$ order)}
	\begin{tabular}{|c|ccc|}\hline
		\diaghead{\theadfont Diag ColumnmnHead II}%
		{$\alpha_k \alpha_l\alpha_m$\\ $\alpha_n\alpha_p\alpha_q\alpha_r\alpha_s$}{\\$\beta_i\beta_j$}&\thead{$\beta_1^2$}&\thead{$\beta_2^2$}& 	 \thead{$\beta_1\beta_2$}  (=$\beta_2\beta_1$)\\
		\hline
		& & &\\
		$\alpha_1^8$& $a_{1111111111}$ & $a_{1111111122}$ &  \\
		$\alpha_2^8$ & $a_{2222222211}$ & $a_{2222222222}$ & \\
		$\alpha_1^2\alpha_2^6$ (x28)& $a_{1122222211}$ &$a_{1122222222}$  &   \\
		$\alpha_1^4\alpha_2^4$ (x70)  &$a_{1111222211}$ & $a_{1111222222}$&  \\
		$\alpha_1^6\alpha_2^2$ (x28) &$a_{1111112211}$ & $a_{1111112222}$&  \\
		$\alpha_1\alpha_2^7$ (x8)& & & $a_{1222222212}$  \\
		$\alpha_1^3\alpha_2^5$ (x56)&  &  &$a_{1112222212}$ \\
		$\alpha_1^5\alpha_2^3$ (x56)&  &  &$a_{1111122212}$ \\
		$\alpha_1^7\alpha_2$ (x8)&  &  &$a_{1111111212}$ \\
				\hline
	\end{tabular}
\end{table}



%

\end{document}